# Combining feature-based approaches with graph neural networks and symbolic regression for synergistic performance and interpretability

Rogério Almeida Gouvêa[1,2,*], Pierre-Paul De Breuck[2,†], Tatiane Pretto[1],

Gian-Marco Rignanese[2,*], Marcos José Leite Santos[1]

[1] Laboratory of Applied Materials and Interfaces, Federal University of Rio Grande do Sul, Porto Alegre, RS 91501-970, Brazil.

[2] Institute of Condensed Matter and Nanosciences, Université Catholique de Louvain, Louvain-la-Neuve, Belgium.

[†]Current address: Ruhr-Universität Bochum, Universitätsstr. 150, 44801 Bochum, Germany.

*Corresponding authors: rogeriog.em@gmail.com; gian-marco.rignanese@uclouvain.be;

**ABSTRACT**

This study introduces MatterVial, an innovative hybrid framework for feature-based machine learning in materials science. MatterVial expands the feature space by integrating latent representations from a diverse suite of pretrained graph neural network (GNN) models—including structure-based (MEGNet), composition-based (ROOST), and equivariant (ORB) graph networks—with computationally efficient, GNN-approximated descriptors and novel features from symbolic regression. Our approach combines the chemical transparency of traditional feature-based models with the predictive power of deep learning architectures. When augmenting the feature-based model MODNet on Matbench tasks, this method yields significant error reductions and elevates its performance to be competitive with, and in several cases superior to, state-of-the-art end-to-end GNNs, with accuracy increases exceeding 40% for multiple tasks. An integrated interpretability module, employing surrogate models and symbolic regression, decodes the latent GNN-derived descriptors into explicit, physically meaningful formulas. This unified framework advances materials informatics by providing a high-performance, transparent tool that aligns with the principles of explainable AI, paving the way for more targeted and autonomous materials discovery.

**Keywords:** Feature-based machine learning, MODNet, graph neural networks, materials informatics, interpretability.

# Introduction

Machine learning has revolutionized materials science, accelerating material discovery and property optimization across various domains[1–3]. The two prominent approaches in this field are feature-based and graph-neural-network (GNN) models, each with distinct advantages and limitations[4,5]. Feature-based models rely on predefined descriptors such as elemental properties, geometric features, and electronic structure information. They are highly interpretable and effective with small datasets, offering insights into structure-property relationships[6,7]. These models adapt well to custom tasks in experimental settings, such as nanocrystal research[8], catalysis[9], and organic photovoltaics[10]. In contrast, GNN models represent materials as graphs, capturing structural information through message passing and learning deep representations with simple atomic descriptors. This often results in more accurate predictions for complex materials, but requires greater computational resources and data for training[11,12]. GNNs are particularly effective in the large-scale screening of materials and for constructing interatomic potentials owing to their efficient computation and local information aggregation,[13] however they lack interpretability.

Boosting the accuracy of feature-based models to make them competitive on larger datasets usually implies employing neural network models and relying on extensive suites, such as MatMiner[7], to produce meaningful features. This process is particularly time-consuming for sophisticated descriptors like the Orbital Field Matrix (OFM)[14] and the Smooth Overlap of Atomic Positions (SOAP)[15]. A novel strategy to boost these feature-based models involves leveraging the rich latent-space representations learned by GNN models pretrained on vast datasets. Even though neural networks are universal function approximators, easing their burden through well-aligned feature transformations can improve generalization, reduce training time, and stabilize convergence[16,17].

In this work, we address these challenges by proposing a hybrid approach that combines traditional chemically intuitive descriptors with latent features obtained from a diverse set of pretrained models. We incorporate features from both structure-based (MEGNet, coGN)[18,19] and composition-based (ROOST)[20] GNNs, as well as from ORB[21], a powerful equivariant Machine Learning Interatomic Potential (MLIP). To avoid the featurization bottleneck of traditional descriptors, we also leverage GNNs to generate fast, latent-space approximations of MatMiner ($\ell$-MM) and Orbital Field Matrix ($\ell$-OFM) features. Finally, we augment this feature set with new descriptors derived via symbolic regression. This multifaceted strategy aims to create a more robust, accurate, and versatile featurizer that capitalizes on the distinct strengths of each approach to be useful for a wider range of dataset sizes.

To simplify the generation of all those features, a package was developed named MatterVial standing for **MAT**erials fea**TuR**e **E**xtraction **V**ia **I**nterpretable **A**rtificial **L**earning, which, besides producing all latent-space features from the GNN models, aids in obtaining the interpretable chemical descriptors that correlate to these high-level features. This is achieved through techniques such as SHapley Additive exPlanations (SHAP) analysis in surrogate models and symbolic regression via Sure Independence Screening and Sparsifying Operator (SISSO) to obtain an approximate formula from the most important features. Our results demonstrate an overall improvement in all analyzed datasets compared with the baseline MatMiner featurizer. In addition, it surpassed the performance of the individual GNN models in several cases, indicating that the combination of traditional and latent-space features leads to a more robust generalization.

This work is situated within a recent methodological trend that repurposes GNNs not as end-to-end predictors, but as powerful and data-efficient feature generators for a variety of downstream tasks[22–26]. Our approach bridges feature-based and graph-based methods, leveraging their strengths to develop more versatile and task-agnostic machine learning models in materials science. By enhancing the accuracy, efficiency, and interpretability of property prediction, this framework facilitates the integration of both experimental and simulated data. Moreover, it aligns with the growing demand for explainable AI[27,28], which is essential for the advancement of self-driving laboratories in materials discovery and optimization[29].

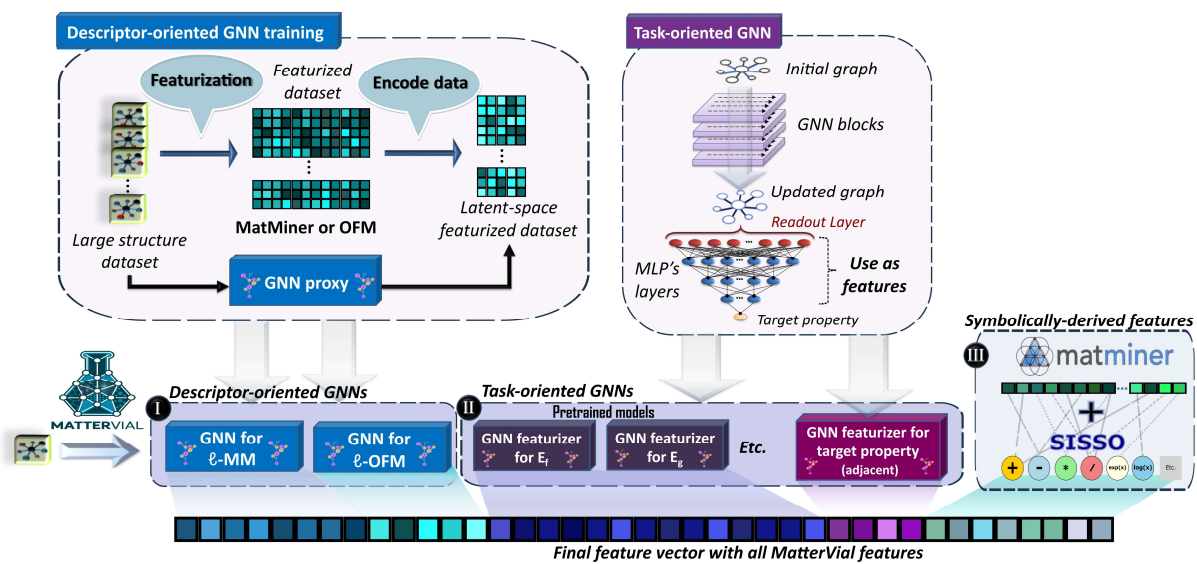

**Fig. 1** | Overview of the methodology for leveraging latent-space features from GNN models with MatterVial. On the left (I), the generation and deployment of descriptor-oriented GNN models are illustrated. At the center (II), task-oriented GNN models—either pretrained or trained on the fly for adjacent variants—are shown, with feature extraction possible from activation, pooling, or multi-layer perceptron (MLP) layers on the model architecture. On the right (III), formulas from symbolic regression from SISSO are also implemented, leveraging traditional physiochemical descriptors available in MatMiner as a base.

## Results and discussion

We evaluate our approach using the full MatBench v0.1 benchmark[30] with MODNet[4], which is the state-of-the-art feature-based model in materials science[12]. We adopt the same MatMiner featurization as that used in MODNet for MatBench in the original publication[4,31]. These can be complemented by three categories of MatterVial features, as illustrated in Fig. *1*:

I. **Latent-space features from descriptor-oriented GNNs:** Conventional material descriptors are transformed into latent representations using an autoencoder trained on Materials Project (MP) data. These descriptors include the widely used MatMiner features ($\ell$-MM) and the features from the Orbital Field Matrix featurizer ($\ell$-OFM). A GNN was then trained to replicate these latent features directly from the input structures. This method achieves a computational efficiency similar to that of GNNs and still preserves interpretability via decoding.

II. **Latent-space features from task-oriented GNNs:** These features are extracted directly from the intermediate layers of pretrained GNN models that have been developed for various tasks. Specifically, we incorporate MEGNet models from the Materials Virtual Lab (MVL) that were pretrained for the prediction of elastic constants, band gap, and formation energy, as well as for the metal-insulator classification. We also consider composition-based ROOST models for the band gap and formation energy. In addition, we include the internal layers of ORB-v3, a state-of-the-art equivariant MLIP trained to reproduce energies and forces. This group capitalizes on the strengths of GNN architectures in capturing complex structural representations, aiming to enhance predictive performance on larger datasets.

III. **Symbolically-Derived Feature Combinations:** Here, we use the MatMiner features as a basis to generate new compound features. Through symbolic regression with SISSO, we identify several combinations of pairs of features (rung one) that exhibit enhanced correlations with the target properties of interest in materials science. These derived formulas are then incorporated as new features.

Since the features obtained from task-oriented GNNs are high-level and not directly interpretable as traditional descriptors, we develop a method to decompose them into interpretable descriptors, which is integrated in the Interpreter module in MatterVial. In addition, features from descriptor-oriented GNNs can be decoded in their interpretable counterparts. Equally, the third group of augmented features via symbolic regression can have their formulas retrieved by name. Comprehensive implementation details for each category and for the Interpreter module are available in the Methods section and in the *Supplementary Information*.

## MatBench validation of MatterVial features

Table 1 presents the performance of MODNet using MatMiner augmented with MatterVial descriptors (MODNet@MM+MV) and MODNet using only MatterVial descriptors (MODNet@MV) relative to the baseline model using only MatMiner features (MODNet@MM) in the 13 MatBench tasks. The results show that blending both latent-space representations from task-oriented and descriptor-oriented GNNs with symbolically derived features consistently reduces prediction errors across this diverse array of property prediction tasks.

Our approach significantly improves the performance on smaller datasets, where feature-based models have traditionally outperformed GNNs. Specifically, our models set new performance records for four tasks previously led by MODNet@MM and now achieve a leading performance in metallicity classification from experimental data. Notably, the glass-forming ability task alone did not result in substantial improvements. We highlight that for smaller composition-based datasets, MatMiner featurization is sufficiently fast to make MODNet@MM+MV computationally effective. For larger datasets, in which traditional featurization is very time consuming, our MODNet@MV models significantly bridge the gap between feature-based and graph-based models, even outperforming state-of-the-art (SOTA) models in predicting properties such as elastic constants, band gap, metallicity, and formation energy. This success demonstrates that our approach effectively addresses the common shortcomings of both feature- and graph-based models. Note, however, that some of the larger MatBench tasks can no longer be considered truly independent test sets for models exposed to vast amounts of similar ab initio data during pretraining.

**Table 1** | Performance comparison of three MODNet variants against the best multi-purpose MatBench model on each task in the MatBench v0.1 benchmark. Metrics are reported as mean absolute error (MAE) for regression and area under the receiver-operator curve (AUROC) for classification. MODNet@MM uses only MatMiner features; MODNet@MM+MV augments these features with MatterVial descriptors; and MODNet@MV uses only MatterVial features essentially substituting MatMiner features by $\ell$-MM. For each task, the MatterVial feature group that yields the best result is shown. Scores in bold identify the overall best model per task, and shaded tasks are those in which MODNet was already the best model.

| MatBench task | $n$ | MODNet@ MM (baseline) | MODNet@ MM+MV (% error reduction*) | MODNet@ MV (% error reduction*) | MatBench record (model) | Best MatterVial groups** |
|---|---|---|---|---|---|---|
| Steels yield strength (MPa) [32] | 312 | 87.76 | **85.12 (3.0%)** | 120.95 (-37.8%) | MODNet | ROOST |
| $E_{exfol.}$ (meV/atom) [33] | 636 | 33.19 | 29.19 (12.1%) | **28.86 (13.0%)** | MODNet | ORB, MVL, $\ell$-OFM, $\ell$-MM |
| argmax(PhDOS) (cm$^{-1}$) [34] | 1,265 | 34.27 | 30.08 (12.2%) | 30.58 (10.8%) | **28.76** (MegNet) | MVL, ORB, $\ell$-OFM, ROOST, SISSO |
| Exp. band gap (eV) [35] | 4,604 | 0.333 | **0.290 (12.9%)** | 0.351 (-5.5%) | MODNet | ROOST, SISSO |
| Refractive index [36,37] | 4,764 | 0.271 | 0.235 (13.3%) | **0.234 (13.7%)** | MODNet | ORB, $\ell$-OFM, MVL, $\ell$-MM |
| Exp. metallicity (eV) [35] | 4,921 | 0.916 | **0.976 (71.4%)** | 0.898 (-59.3%) | 0.921 (AMMExpress) | ROOST |
| Glass-forming ability [38,39] | 5,680 | 0.936 (**0.960**)† | 0.937 (1.6%) | 0.904 (-50.0%) | MODNet | ROOST |
| Logarithmic $G^{vrh}$ (log$_{10}$GPa) [40] | 10,987 | 0.073 | **0.032 (55.5%)** | 0.033 (54.8%) | 0.067 (coGN) | MVL, ORB, $\ell$-MM, ROOST, SISSO |
| Logarithmic $K^{vrh}$ (log$_{10}$GPa) [40] | 10,987 | 0.056 | **0.027 (49.6%)** | 0.028 (50.1%) | 0.049 (coNGN) | MVL, ORB, $\ell$-MM, $\ell$-OFM, ROOST, SISSO |
| Perovskite $\Delta H_{form}$ (eV/unitcell) [41] | 18,928 | 0.0908 | 0.0386 (57.5%) | 0.0389 (57.3%) | **0.0269** (coGN) | ORB, MVL, $\ell$-OFM, $\ell$-MM, ROOST, SISSO |
| Band gap (eV) [37] | 106,113 | 0.2199 | 0.137 (37.6%) | **0.137 (37.8%)** | 0.156 (coGN) | MVL, ORB, ROOST, SISSO |
| Metallicity [37] | 106,113 | 0.904 | **0.978 (77.1%)** | 0.976 (75.0%) | 0.9520 (CGCNN) | ORB, MVL, $\ell$-OFM, ROOST |
| $E_f$ (eV/atom) [37] | 132,752 | 0.0448 | 0.0147 (67.2%) | **0.0138 (69.2%)** | 0.0170 (coGN) | MVL, ORB, $\ell$-OFM, $\ell$-MM |

* % error reduction $= \frac{MAE_{baseline} - MAE_{model}}{MAE_{baseline}} \times 100\%$ (regression) $or$ $\frac{(1-AUROC_{baseline}) - (1-AUROC_{model})}{(1-AUROC_{baseline})} \times 100\%$ (classification)

** Ordered by importance, MVL, ORB and ROOST refer to the task-oriented GNN features, respectively those from MVL MEGNet models for structures, the MLIP Orb-v3 and pretrained ROOST models for compositions. $\ell$-MM and $\ell$-OFM refer to the descriptor-oriented GNN features, $\ell$-MM when included, substitutes the MatMiner features for faster generation. SISSO refers to the group of features derived from MM features via symbolic regression.

† As we were unable to replicate the reported 0.960 AUROC for glass formability using MODNet, we present our best MODNet@MM result as baseline instead. Despite the lower score, MODNet continues to outperform other models in MatBench for this task.

An analysis of the feature contributions in Table 1 reveals that task-oriented latent features are the primary drivers of performance gains. The inclusion of ROOST aimed at enhancing performance in composition-based tasks, and yet the model has reliably improved results in a wide range of tasks that also contained structural information. This performance may be attributed to the attention mechanism that captures unique patterns during activation and material pooling. For structure-based tasks, MVL-derived features have shown a significant positive impact. They boost predictions even when the prediction targets differ from those used in the original models, such as in predicting the perovskite heat of formation and refractive index. The ORB features, derived from an equivariant MLIP, proved particularly impactful, frequently appearing as top contributors. This is chemically intuitive, as the model's training on energies and forces provides a rich, physically meaningful latent space that is useful for transfer learning. This aligns with very recent findings by Kim et al.[26], who also employed ORB features with MODNet for structure-based regression tasks. Our approach achieves enhanced performance by incorporating all Orb-v3 layers and combining these features with diverse descriptor groups within our framework.

The descriptor-oriented and symbolically derived features also provided consistent complementary improvements. The $\ell$-OFM features improved performance across most tasks, validating that our GNN-based approximation is an efficient and effective method for incorporating the descriptive power of computationally expensive descriptors like the Orbital Field Matrix. The $\ell$-MM features, designed as a shortcut for MatMiner features via GNN, lead to improved or similar performance on many tasks. Compared to the models that used the full MatMiner features (MODNet@MM+MV), we argue that the reconstruction loss was sufficiently low and that, for some cases, the encoder effectively refined the representation via regularization, improving the metrics. Crucially, these latent-space representations remain decodable, preserving much of the interpretability, which is a hallmark of feature-based models. Finally, the SISSO-derived features, while less universally impactful, still boosted performance in roughly half of the benchmarks. Given that we utilized only first-rung symbolic regression, we conjecture that there is clear potential for further gains with higher-level, more complex formulas. Ultimately, these results show that our approach simultaneously accelerates featurization, improves model performance, and provides valuable chemical insights. This combination of benefits repositions feature-based models as strong and practical alternatives to end-to-end GNNs for property prediction.

**Synergy of MatterVial features and adjacent GNN model**

Having demonstrated the performance gains of our method, we now turn to the individual contributions of the MatterVial features. We examine the synergetic effects of each MatterVial feature group using the perovskite heat of formation task as an example. Table 2 illustrates a step-by-step performance evaluation for this task, revealing how the integration of different MatterVial feature groups leads to cumulative improvements. Starting from our baseline, the MODNet@MM model delivers an MAE of 0.0888 eV/unit cell. This performance serves as a reference point against which the benefits of the additional features can be measured.

The first modification involves introducing descriptor-oriented GNN features, $\ell$-OFM and $\ell$-MM, which are designed to be computationally faster approximations of their full counterparts. When MatMiner features are entirely replaced by their latent representation (MODNet@$\ell$-MM), the MAE is 0.1052 eV/unit cell. While higher than our MODNet@MM baseline (0.0888 eV/unit cell), this still significantly outperforms AutoMatMiner (0.2005 eV/unit cell), demonstrating $\ell$-MM as a viable, faster featurization alternative. Augmenting MatMiner with $\ell$-OFM (MODNet@MM+$\ell$-OFM) reduces the MAE to 0.0794 eV/unit cell. This is lower than the baseline, although still higher than that obtained using the original computationally intensive OFM features (MODNet@MM+OFM, 0.0751 eV/unit cell). Combining both $\ell$-MM and $\ell$-OFM (MODNet@$\ell$-MM+$\ell$-OFM) yields an MAE of 0.0973 eV/unit cell. These results highlight that our proxy GNN featurizers offer a compelling trade-off, capturing essential chemical information with a substantial speed-up in featurization.

Building on this foundation, the incorporation of task-oriented GNN features from the MVL pretrained models further boosts performance in MODNet@ℓ-MM+ℓ-OFM+MVL model, lowering the MAE to 0.0673 eV/unit cell. Clearly, the MVL descriptors capture additional structural and physicochemical details that the MM and OFM features do not, thereby enhancing the ability of the model to predict heat formation (more details on the MVL descriptions and the effect of different layers are given in the *Supplementary Information, section S6*).

Next, the addition of symbolically derived feature combinations via SISSO produces modest refinement, reducing the MAE to 0.0653 eV/unit cell. Although the improvement is small, it underscores the notion that simple algebraic combinations of conventional descriptors can reveal non-linear relationships, complement the latent-space features, and thereby enhance prediction accuracy.

Further refinement is achieved by incorporating composition-based ROOST features. At first glance, one might not expect an improvement over the MEGNet MVL models since they incorporate structural information alongside composition. However, we believe that the attention-based mechanism present in ROOST is responsible for capturing additional meaningful information to complement other feature groups and achieve an MAE of 0.0639 eV/unit cell. Furthermore, at this point, using the standard MatMiner features instead of their latent representation (ℓ-MM) yields a nearly equivalent performance (MAE of 0.0637 eV/unit cell). These results confirm that the rapidly generated encoded representations can effectively replace the full MatMiner features in tandem with other descriptors. However, eliminating MatMiner features entirely (neither MM nor ℓ-MM), causes a significant decrease in accuracy with 0.0707 eV/unit cell in MODNet@MVL+ROOST and 0.0716 eV/unit cell in

MODNet@MVL, indicating that the MatMiner features are valuable and not simply redundant to these GNN descriptors. In fact, a synergistic effect among all MatterVial feature groups is observed in this dataset.

ORB features stand apart from other featurizers like MVL and ROOST. While MVL and ROOST were trained on smaller datasets, specifically MP and OQMD[42] (about 1.5 million structures combined), the ORB-v3 featurizer was trained on a significantly larger dataset. This dataset, which combines MP, Alexandria[43], and OMat[44], leverages approximately 120 million calculated structures, a number at least two orders of magnitude larger than either those datasets. The extraction of features from this model in MatterVial to use in MODNet significantly reduces the mean absolute error in the task, but a slight improvement is still seen with the other MatterVial features that were included. We conjecture that larger reductions might still be achievable by training more task- and descriptor-oriented models in these larger datasets.

Despite the significant reduction, feature-based approaches using pretrained models with MatterVial or HackNIP[26] still fall short of the results obtained purely with GNNs such as MEGNet and coGN trained in the perovskites dataset. Based on this observation, we incorporate into MatterVial the possibility of training adjacent GNN models on the fly and extracting their features with the AdjacentGNNFeaturizer class. We achieve 0.0343 eV/unit cell using the MEGNet adjacent model features. The MEGNet benchmarked MAE is substantially lower than what we achieved using the default configuration of the model, even with the same elemental embeddings provided by the authors. This discrepancy is possibly due to differences in hyperparameters, inclusion of additional features, and larger training times employed for the benchmark[45]. Finally, we employed the SOTA coGN model as an adjacent model for feature extraction, and we obtained comparable results to the reported values in

MatBench with this model. Incorporating coGN features in our MODNet model reduced the MAE to 0.0313 eV/unit cell, which is much closer to the 0.0269 eV/unit cell record.

**Table 2 | Mean absolute errors (MAEs) for the MatBench task of the heat of formation of perovskites with different models.**

| Reference models | MAE (eV/unit cell) | MatterVial models | MAE (eV/unit cell) |
|---|---|---|---|
| ***Descriptor-oriented*** | | | |
| AutoMatMiner (MatBench*) | 0.2005 (±0.0085) | MODNet@$\ell$-MM | 0.1052 (±0.0022) |
| MODNet@MM (this work) | 0.0888 (±0.0025) | MODNet@MM+$\ell$-OFM | 0.0794 (±0.0016) |
| MODNet@MM+OFM | 0.0751 0.0888 (±0.0018) | MODNet@$\ell$-MM+$\ell$-OFM | 0.0973 (±0.0016) |
| ***Task-oriented (MVL, ROOST)*** | | | |
| MODNet@MVL | 0.0716 (±0.0020) | MODNet @$\ell$-MM+$\ell$-OFM+MVL | 0.0673 (±0.0015) |
| MODNet@MVL+ROOST | 0.0707 (±0.0017) | MODNet@$\ell$-MM+$\ell$-OFM+ +MVL+SISSO | 0.0653 (±0.0013) |
| MODNet@MM+$\ell$-OFM+ +MVL+SISSO+ROOST | 0.0637 (±0.001) | MODNet@$\ell$-MM+$\ell$-OFM+ +MVL+SISSO+ROOST | 0.0639 (±0.0010) |
| ***Task-oriented (ORB featurizer)*** | | | |
| MODNet@MM+$\ell$-OFM+ +MVL+SISSO+ROOST+ORB | 0.0386 (±0.0009) | | |
| HackNIP[26] (MODNet@ORB) | 0.0397 | MODNet@MV † | 0.0388 (±0.0006) |
| ***MV + Adjacent GNN model*** | | | |
| MEGNet (MatBench*) | 0.0352 (±0.0016) | | |
| MEGNet (this work) | 0.0685 (±0.0036) | MODNet@ MV+Adj(MEGNet) | 0.0343 (±0.0014) |
| coGN (MatBench*) | 0.0269 (±0.0008) | MODNet@ MV+Adj(coGN) | 0.0313 (±0.0012) |
| coGN (this work) | 0.0271 (±0.0008) | MODNet@ MV+Adj(coGN)+hiSISSO | 0.0288 (±0.0009) |

*Data retrieved from *MatBench*[12] in August 2025.
† For brevity MV = ($\ell$-MM+$\ell$-OFM+MVL+SISSO+ROOST+ORB), i.e. all pretrained featurizers in MatterVial.

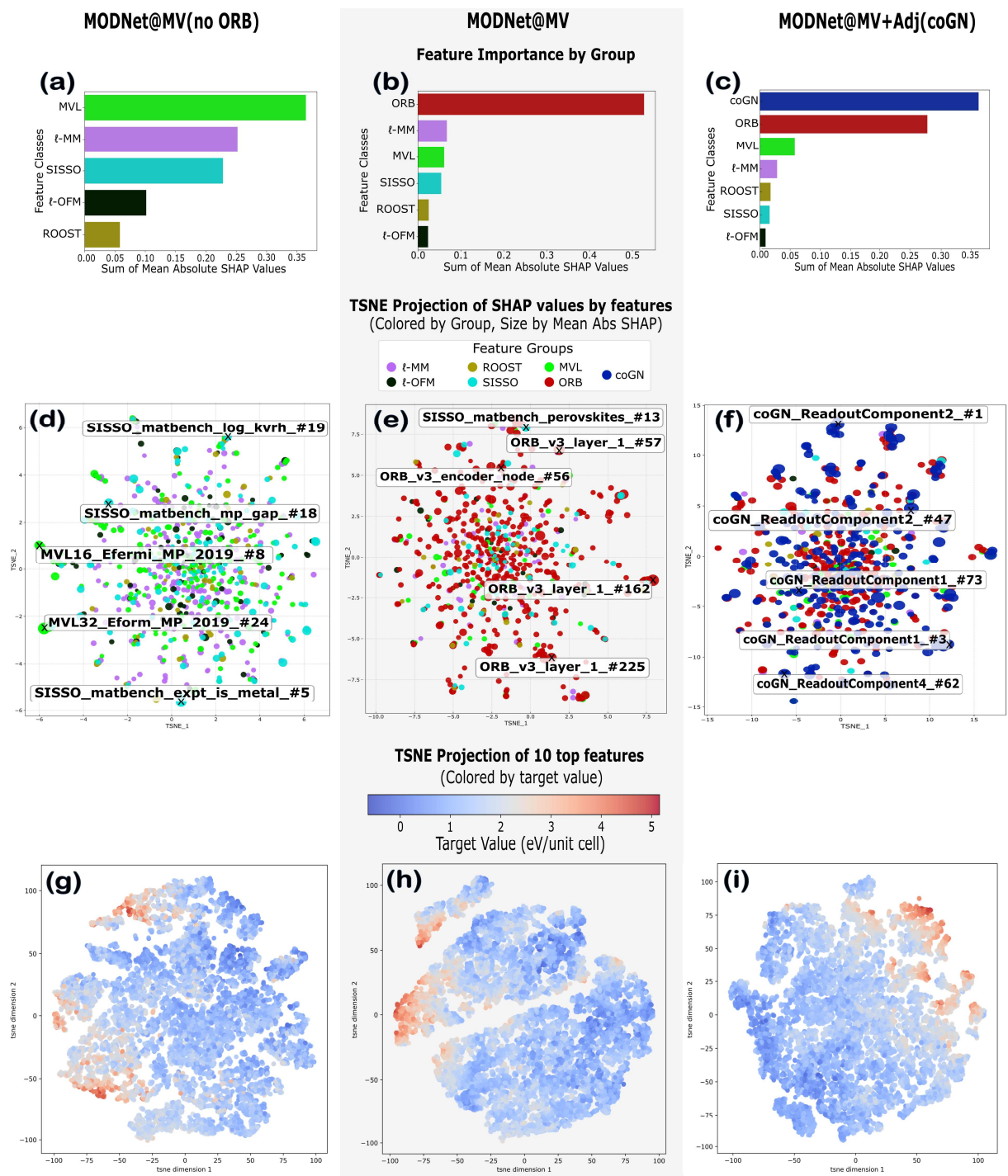

**Fig. 2** | The impact of different MatterVial feature sets on the performance of MatBench's perovskite heat of formation task is illustrated with three different types of plots for each of the three models: on the left, MODNet@MV(no ORB), where features from the ORB model are excluded; at the center, MODNet@MV, which includes all pretrained MatterVial features; and on the right, MODNet@MV+Adj(coGN), which further incorporates features from an adjacent coGN model trained on the task. (a–c) Bar plots showing the feature importance aggregated by a group of features through the sum of the mean absolute SHAP values. (d–f) t-SNE projections of the SHAP values for each feature in the model, colored by the feature group and some of the features with the highest contribution annotated. (g–i) t-SNE projections of the top 10 most important features colored by the target value (heat of formation).

Fig. **2** graphically depicts the synergistic effects detailed in Table 2 by comparing different models. The feature importance from the mean absolute SHAP values aggregated by feature group in Fig. 2(a-c) quantifies the contribution of the different groups of features and shows a clear shift in dominance as more powerful features are introduced. However, a closer inspection reveals important nuances regarding how these feature sets interact. In the MODNet@MV(no ORB) model, there is a relatively balanced and significant contribution from all feature groups, led by MVL, $\ell$-MM, and SISSO, underscoring their collective utility. This is seen more clearly with the t-SNE projections of the SHAP value vectors for each feature in the model, where we can see these three sets of features covering most regions of the projection, but still some contributions of ROOST and $\ell$-OFM features. When ORB features are introduced (Fig. 2b and 2e), they become the dominant contributor, explaining the dramatic reduction in MAE observed in Table 2. Crucially, the $\ell$-MM and SISSO features retain a significant portion of their importance with SISSO, being even among the highest contributors. This indicates that they capture complementary chemical information not fully encapsulated within the ORB latent space, explaining the slightly better result obtained compared to HackNIP's MODNet@ORB model[26]. This hierarchical and synergistic contribution of features directly explains the visual improvement in the data manifold shown in the t-SNE projections (Fig. 2g-i). The feature space of the MODNet@MV(no ORB) model (Fig. 2g) shows some organization. However, the introduction of ORB features (Fig. 2h) creates a significantly more structured manifold with a smoother gradient along the target property.

This synergistic contribution continues in the final MODNet@MV+Adj(coGN) model. The inclusion of adjacent coGN features (Fig. 2i) results in the most well-defined feature space in the t-SNE projection, with the clearest separation between data points according to the target feature. While the task-specific coGN features predictably take the lead, the pretrained ORB and MVL features remain highly influential, serving as the second- and third-most important groups, respectively (Fig. 2c and 2f). In contrast, the contributions from ℓ-MM and SISSO are now marginal, as their predictive information has been superseded by more powerful GNN features. This layered view of contributions highlights the interpretability brought by feature-based models. In the following section, we showcase how this interpretability can be deepened using new MatterVial tools.

**Interpretability of MatterVial features**

We begin by analyzing the most important features of the MODNet@MV(no ORB) model to understand what factors increase its accuracy in predicting the perovskite heat of formation ($\Delta H_f$). Unlike end-to-end GNNs, where features are deeply entangled through message passing, feature-based models have readily decoupled features, and SHAP values can be used to robustly assess the most important ones, as shown in the plot in Fig. **3.** Utilizing the MatterVial Interpreter module, we can easily obtain SISSO formulas with up to five terms to approximate the GNN features of the included pretrained models. These approximations are based on interpretable descriptors from MatMiner and OFM. The plot displays the one-term formulas and their corresponding $R^2$ values, demonstrating that even with relatively simple descriptors, these approximation formulas can achieve high $R^2$ values for many meaningful GNN features.

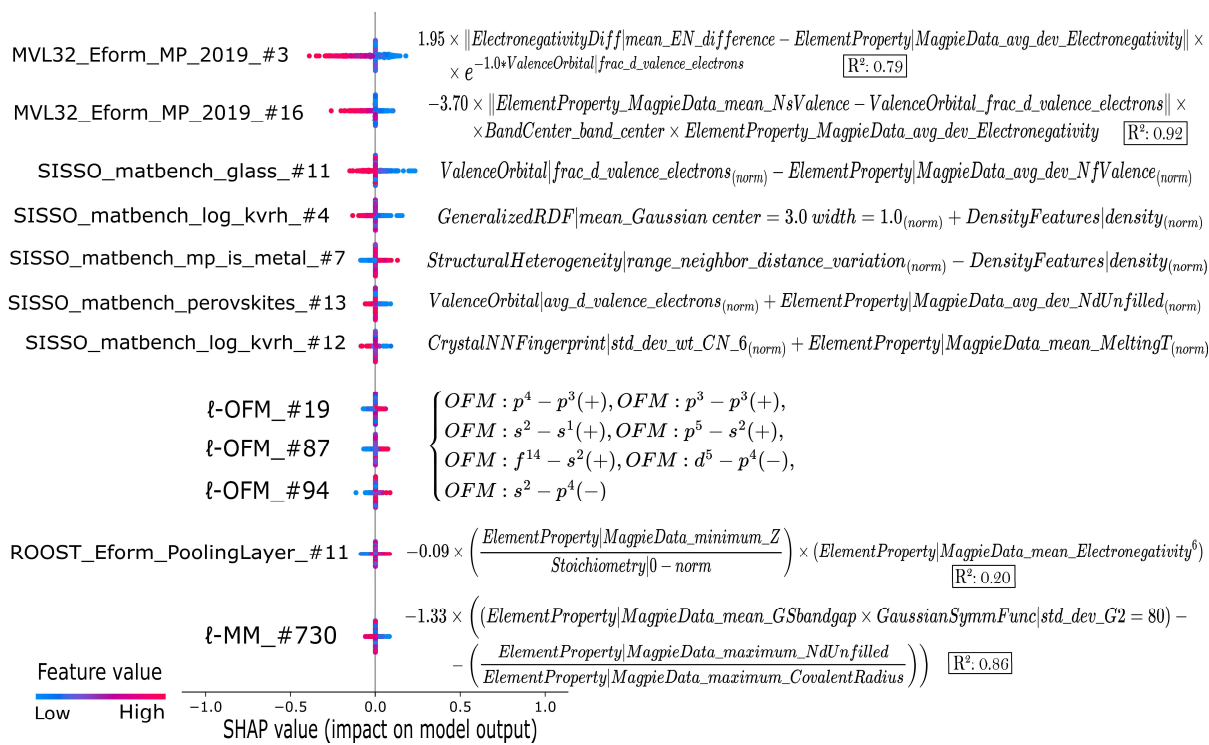

**Fig. 3 | SHAP values plot for selected MatterVial features in the MODNet@MV(no ORB) model for the perovskite heat of formation.** The plot displays the impact of individual features on the model's output (SHAP value), with the color indicating the feature's value (blue for low, red for high). Alongside each feature, the corresponding 1-term SISSO formula approximation for the MatterVial features and its R² value, when appropriate, are shown.

This analysis identifies several key feature groups that drive the predictions. Features from the MVL formation energy model, for instance, correlate stability with large electronegativity gaps (promoting ionic character) and low d-electron fractions, which favor early transition metals. Features generated by SISSO highlight structural drivers, rewarding dense atomic packing, ordered coordination environments, and specific stabilizing factors like 3Å interatomic contacts, while penalizing destabilizing electronic effects from excess d-electrons. Compositional features from ROOST and encoded MatMiner ($\ell$-MM) models capture broader trends, showing that perovskites made of heavier, chemically diverse elements tend to be less stable and illustrating the balance between destabilizing wide-band-gap elements and the stabilizing effect of species with many unfilled d-states. Finally, encoded OFM ($\ell$-OFM) features provide a granular view of bonding, distinguishing between the stabilizing interactions characteristic of oxides (e.g., $s^2$–$p^4$) and weaker bonds involving pnictogens. Collectively, this demonstrates that the model learns a multi-faceted and physically grounded understanding of perovskite stability. A full breakdown of the individual features shown in the figure is provided in the *Supplementary Information,* section S10.

A comparative SHAP analysis of the best-performing MODNet@MV and MODNet@MV+adj(coGN) models, which incorporate richer ORB and coGN features (see SI, section S11, Figs. S4-S7), showed that while the MODNet@MV(no ORB) model primarily relies on fundamental chemical descriptors, the addition of ORB features shifts the emphasis of the model toward geometric information such as packing efficiency. The top-performing MODNet@MV+adj(coGN) model builds on this by capturing the most sophisticated features, representing a complex interplay between chemical and geometric properties. This increase in predictive power is accompanied by a decrease in direct interpretability. As the models become more complex, the ability to approximate their most important features with simple SISSO formulas diminishes (indicated by progressively lower $R^2$ values), and their correlation with classical descriptors weakens (Table S13). This progression highlights the gap between the complex features of high-performing GNNs and the limited descriptive power of interpretable descriptors, emphasizing the need for more flexible descriptors that remain compact for symbolic regression methods and interpretability.

To test the utility of our GNN feature approximations, we conducted a two-stage experiment. In the first stage, we compared two types of SISSO models: a baseline using only MatMiner and OFM descriptors, and an enhanced version that added formulas approximating the GNN's most important features. For both model types, we apply a consistent methodology, utilizing several primary feature pre-selection algorithms—including mRMR (i-SISSO)[46], random forest importances (rf-SISSO)[47], and our xgb-rfe-SISSO (SI, Sec. S8). The addition of the GNN-derived features yields a significant and consistent reduction in prediction error, as shown in Fig. 4(a). Our approach is analogous to hierarchical SISSO (hiSISSO)[48], but it uniquely feeds back approximations of learned GNN features rather than terms from a prior SISSO model. In the second stage, we extract the terms from this enhanced SISSO model and incorporate them as new "hiSISSO features" to augment the MODNet@MV+adj(coGN) model. This augmented model further reduces the error to 0.0288 eV/unit cell. The t-SNE projection of SHAP value contributions and average feature importance of the classes in Fig. 4(b,c) confirm their effectiveness, showing the high per-feature predictive power of hiSISSO features complementing the model. This demonstrates that explicit, interpretable formulas can improve generalization and raises the compelling question of whether GNN features could be replaced entirely if more expressive, physically grounded descriptors were available.

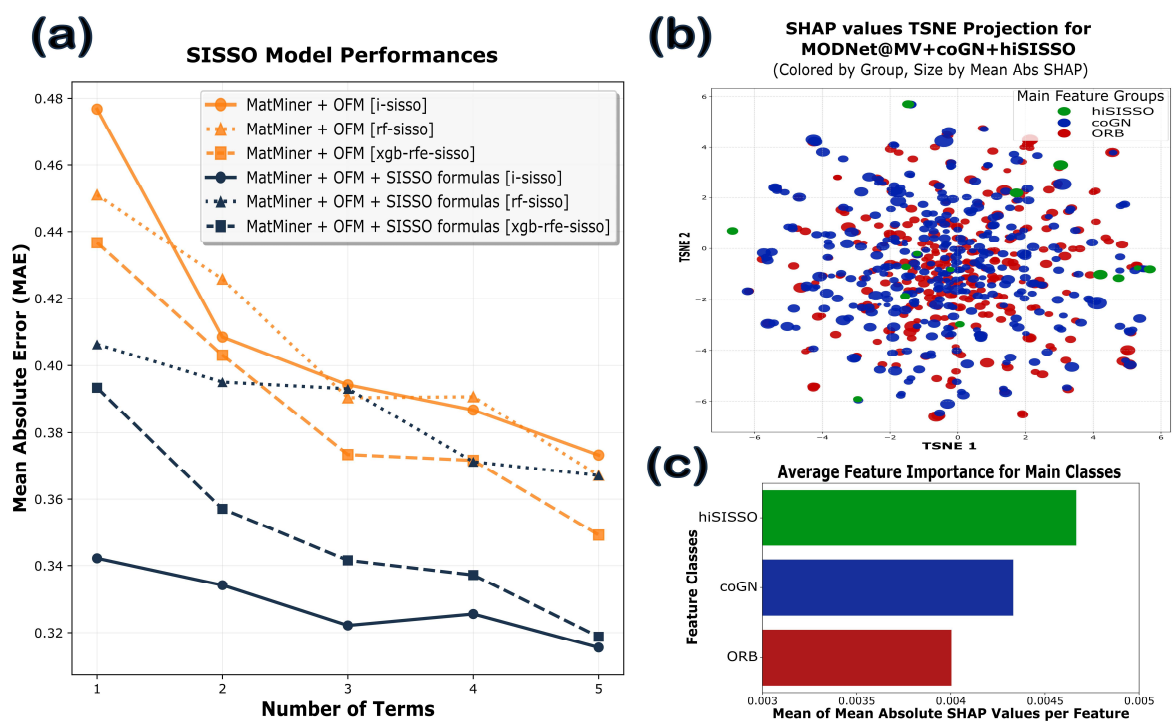

**Fig. 4** | SISSO models and hiSISSO-enhanced MODNet model analysis on matbench_perovskites task. (a) Mean absolute error of models with baseline MatMiner+OFM features (orange) vs. those augmented with SISSO formulas approximating the best GNN features (dark green). Feature selection methods include i-SISSO, rf-SISSO, and xgb-rfe-SISSO; (b) t-SNE projection of SHAP values for top feature groups in the final MODNet@MV+adj(coGN)+hiSISSO model; point size reflects feature impact. (c) Average feature importance across main classes in the final model, calculated from the mean absolute SHAP values.

# Conclusion

In this work, we introduced MatterVial, a unified and modular hybrid framework designed to bridge the gap between the predictive power of graph neural networks (GNNs) and the chemical transparency of traditional feature-based models in materials science. By augmenting the state-of-the-art feature-based model MODNet with a diverse and synergistic set of descriptors, this approach elevates its performance to be competitive with, and in several cases superior to, end-to-end GNNs. To summarize our contributions:

(i) MatterVial is a novel open-source Python framework that generates a rich hybrid feature set. It integrates latent-space representations from various pretrained models, including structure-based GNNs such as MEGNet, an equivariant interatomic potential (ORB), and composition-based networks such as ROOST. The framework also uses computationally efficient GNN-approximated descriptors ($\ell$-MM, $\ell$-OFM) and features derived from symbolic regression.

(ii) The hybrid model demonstrates broad applicability and superior performance across the full MatBench v0.1 benchmark. It consistently reduces prediction errors across nearly all 13 tasks and establishes new state-of-the-art records for feature-based models in several categories.

(iii) A key innovation is a method that systematically decodes abstract GNN-derived features into more intuitive formulaic descriptors. This is achieved using surrogate models and symbolic regression to translate latent representations into explicit mathematical expressions based on fundamental physicochemical properties.

(iv) By incorporating features from an adjacent, task-specific GNN model, the framework enables a feature-based model to achieve predictive accuracy that is highly competitive with state-of-the-art GNNs while uniquely maintaining a modular and analyzable feature space.

(v) It was demonstrated that the interpretable formulas extracted from GNNs can be fed back into the model as new "hiSISSO features", leading to a further reduction in prediction error. This confirms that the interpretability method can capture causally relevant physical information.

In conclusion, this work repositions feature-based modeling as a premier methodology in materials informatics. It delivers a practical solution that meets the dual demands of high accuracy and interpretability, a combination that is becoming increasingly critical in the field. While predictive accuracy is essential, interpretability allows researchers to validate that models have learned physically meaningful principles, thereby building trust and moving beyond simple prediction to genuine scientific understanding. This deeper insight accelerates materials discovery by enabling a shift from brute-force screening to more targeted, hypothesis-driven design. Ultimately, this alignment with the principles of explainable AI is a prerequisite for developing the next generation of autonomous discovery platforms, or "self-driving labs", which require models that can not only predict outcomes but also explain the underlying principles to guide subsequent experiments.

# Methods
## MODNet model training

The MatMiner featurizer used throughout this work is described in detail in the *Supplementary Information,* section S1. For all experiments incorporating MatterVial features, since many features are obtained, we perform an initial preselection of features using recursive feature elimination with XGBoost[49] to reduce the pool to 800 features. Subsequently, the built-in MODNet feature selection algorithm is used to select and rank a subset of these features that will be used for training. At this point, we can determine which groups of MatterVial features are relevant for a given task ("Best MatterVial groups" in Table 1). The MODNet models are optimized via a genetic algorithm to select the best hyperparameters, and the optimal models in the validation set form deep ensembles, as described in Ref. 26, which are then used for evaluation in the test set and to obtain the final metrics.

The mean absolute error (MAE) serves as the primary evaluation metric in regression tasks, and for classification tasks, the area under the receiver-operator curve (AUROC) is used. We consistently use a five-fold cross-validation method, as described in Matbench[25,] in all presented tasks. A Supplementary data repository with detailed results of our work is available at https://github.com/rogeriog/MatterVial_SupportData.

## MatterVial implementation

**MatterVial** is an open-source featurizer tool implemented in Python (available at https://github.com/rogeriog/MatterVial) to enhance material property predictions by integrating pretrained descriptor-oriented and task-oriented GNNs, as well as precomputed symbolic formulas from traditional chemically intuitive descriptors. The package offers significant flexibility and modularity, allowing the extraction of features from different layers of pretrained models and the incorporation of other GNN models as needed. The following outlines each MatterVial featurizer employed:

● **ℓ-OFM featurizer:** the OFM featurizer captures valence electron interactions at each atomic site by employing a weighted vector outer product of one-hot encoded valence orbitals for every atom (details in the *Supplementary Information,* section S2, Fig. S1). The structural representation is achieved by averaging all local OFMs. We apply the OFM featurizer to a subset of the Materials Project MP-crystals-2018.6.1[50] dataset with 106,113 structures whose energy above the convex hull was lower than 150 meV, nicknamed MP2018-stable, followed by training an autoencoder to derive a latent space representation. The latent OFM features are subsequently used as targets to train a GNN model that generates these features directly from the initial structures.

● ℓ-MM featurizer : following a similar procedure to the OFM featurizer, we encode features obtained from the default MatMiner featurizer of MODNet v.0.1.13 applied to the MP2018-stable dataset, resulting in 1,336 MatMiner features. The selected compression level provides latent MatMiner features (ℓ-MM), which are then used as targets to train a GNN model that directly generates these features from the original structures.

The **DescriptorMEGNetFeaturizer** class in the MatterVial package is implemented to retrieve OFM-encoded and MatMiner-encoded features from the MatterVial package. A thorough investigation of these encoded features, including the use of different compression levels and hyperparameters was conducted, as detailed in the *Supporting Information* (sections S4, S5, S7 and Fig. S3, also Tables S1–S3, S5–S8, S10–S11).

● **MVL MatterVial featurizers:** Utilizing the **MVLFeaturizer** class from the MatterVial package, we incorporate five pretrained MEGNet models provided by the Materials Virtual Lab[50]. Specifically, these are the models trained for the formation energy, Fermi energy, and elastic constants $K^{VRH}$ and $G^{VRH}$ on the 2019.4.1 Materials Project crystals dataset, as well as the band gap regression model trained on the 2018.6.1 Materials Project crystals dataset. The default MEGNet architecture comprises MEGNet blocks followed by an MLP with two dense layers, one with 32 neurons and the other with 16 neurons, before producing the target property (see section S3, Fig. S2, in *Supplementary Information*). The modularity of the MatterVial package allows us to extract features from different layers of these pretrained models. We extract features from the MLP layers preceding the output, specifically from the 32-neuron (layer32) and 16-neuron (layer16) configurations. An investigation was conducted on the effect of using the different layers for prediction as provided in *Supplementary Information*, section S6, Table S4. For this paper, the extracted features of both layers (160 descriptors for layer32 and 80 descriptors for layer16) are concatenated and added to the final feature vector.

- **Adjacent GNN featurizer:** The **AdjacentGNNFeaturizer** class from the MatterVial package is employed to train a MEGNet or coGN model on the fly for each fold of the train-test split. This adjacent model captures task-specific data nuances, enhancing prediction accuracy. The default hyperparameters from MEGNet v.1.3.2 and coGN are utilized, as detailed in the *Supplementary Information,* section S7.2, Table S9.

- **SISSO-based formula featurizer:** The SISSO++ framework[51] was used to generate symbolic expressions that approximate target material properties across 15 datasets (see *Supplementary Information,* section S8, Table S12 for details) by transforming MatMiner features. The method begins by recursively applying a predefined set of operators (e.g., addition, subtraction, multiplication, division, sine, cosine, exponential, and logarithm) to expand the feature space, followed by sure-independence screening (SIS) that ranks the resulting candidates by their correlation with the target property and a sparsification step that selects a compact descriptor set. For our configuration, restricted to rung one, this yields 20 paired-feature formulas. By opting for the expressions produced at the SIS step instead of the final SISSO formula, versatility and generalization are assured when integrated with MODNet neural networks. These formulas, derived for each of the 15 tasks, are compiled in the file SISSO_FORMULAS_v1.txt, which is accessed by the **get_sisso_features** function in MatterVial to process the given MatMiner features (either directly or decoded from ℓ-MM) and outputs a dataframe of evaluated expressions.

In terms of computational cost, generating the complete feature set with MatterVial is substantially more efficient than traditional MatMiner featurization. Although the precise runtime for MatMiner is highly dataset-dependent, our observations indicate that MatterVial reduces feature generation time by a minimum of two orders of magnitude, especially when leveraging GPUs.

**Retrieving interpretability via MatterVial's interpreter module**

Before employing MatterVial's interpreter module, we conduct a SHAP value analysis (see *Supplementary Information,* section S9) on our MODNet models to assess feature importance. Using the SHAP Python library, we perform the analysis with 300 samples and 500 perturbations on 24 CPU cores in about 20 minutes, revealing the features with the greatest impact on the model predictions.

To bridge the gap between high-level latent representations and interpretable chemical descriptors, MatterVial leverages surrogate XGBoost models. These models are trained to predict each latent feature based on the previous assessment using the MP2018-stable dataset featurized with interpretable MatMiner and OFM features. The tree-based additive structure of XGBoost ensures rapid and parallel training as well as efficient SHAP calculations. For each feature, the top 30 most influential interpretable descriptors, as determined by SHAP, are forwarded to SISSO++, which performs symbolic regression to retrieve a symbolic formula that better correlates with the latent feature. This process is illustrated in Fig. 5. The SHAP decompositions and SISSO formula for the latent-space features computed this way can be retrieved by calling the **Interpreter** class and invoking *get_shap_values* or *get_formula* with the feature name as generated by MatterVial.

Moreover, this interpretability framework extends to adjacent GNN models. Using MatterVial's tools, including the AdjacentGNNFeaturizer, a task-specific GNN model is trained and its latent features can be interpreted following the previous pipeline for which helper functions are provided. In this way, both the pretrained models imported by MatterVial and the adjacent models trained on the fly benefit from enhanced transparency, enabling users to decode the underlying chemical principles driving the predictions.

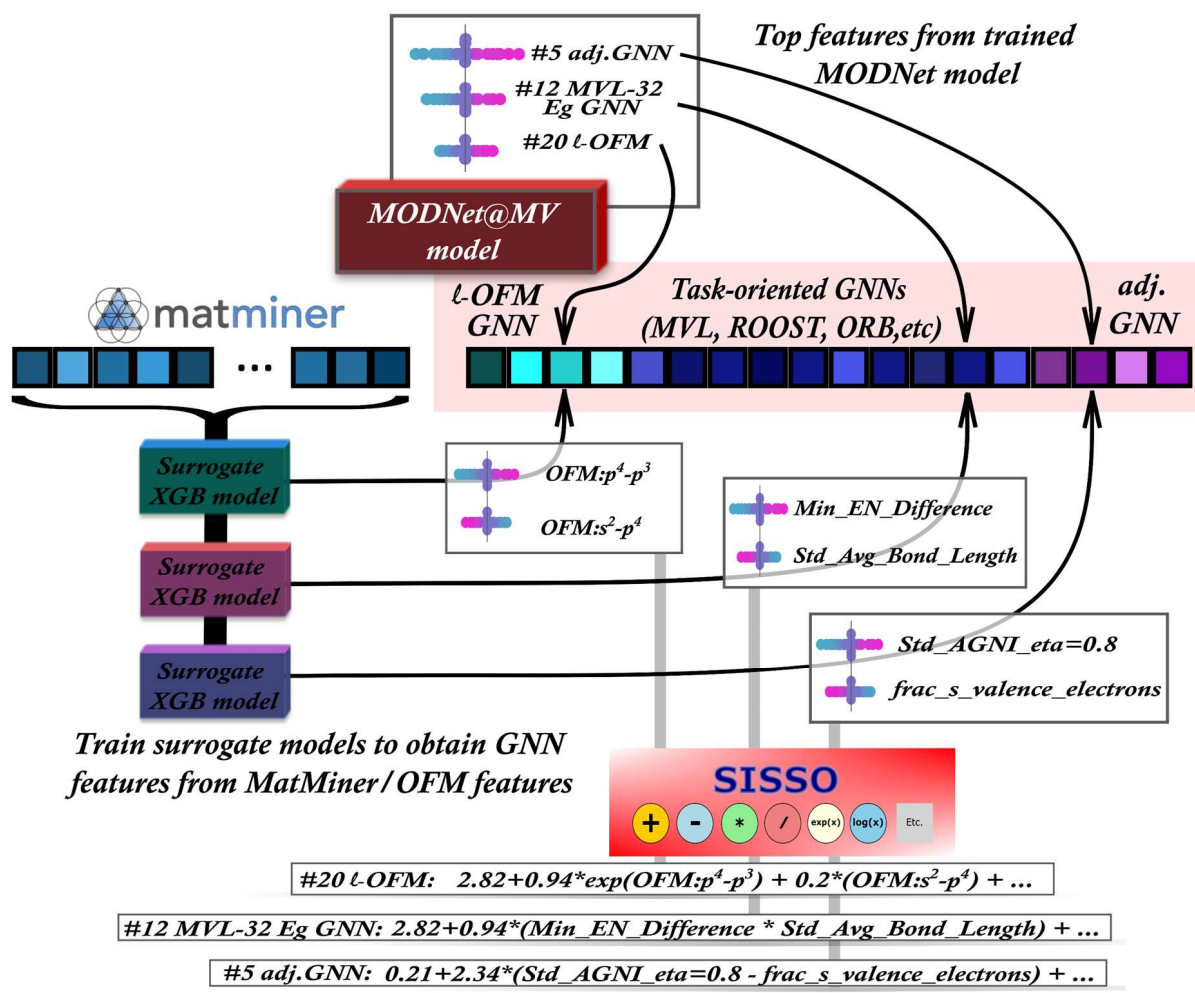

**Fig. 5** | Illustration of the usage of XGBoost models to decompose the most important MatterVial features from the MODNet model into intuitive chemical features via SHAP plots, which can then be used for SISSO symbolic regression.

# Declarations

## Acknowledgements

We acknowledge the supercomputing facilities of the Université catholique de Louvain (CISM/UCL) and the Consortium des Équipements de Calcul Intensif en Fédération Wallonie Bruxelles (CÉCI) for computational resources. This work also made use of Lucia, the Tier-1 supercomputer of the Walloon Region.

**Funding**

This study was financed in part by the Coordenação de Aperfeiçoamento de Pessoal de Nível Superior – Brazil (CAPES) – Finance Code 001 and Université catholique de Louvain (Ref. No. ARH/MKK/01155839).

**Author contributions**

R.G., P.D.B., and G.M.R. conceptualized the study. R.G. performed the experiments and data analysis, and wrote the main manuscript. P.D.B. and G.M.R. supervised the investigation. R.G., P.D.B., T.P., G.M.R. and M.J.L.S. reviewed and approved the final manuscript.

**Competing interests**

The authors declare no competing interests.

# Supplementary Information - Combining feature-based approaches with graph neural networks and symbolic regression for synergistic performance and interpretability

## S1. MatMiner features

The MatMiner features included in this work were implemented in the ***DeBreuck2020Featurizer*** class in the MODNet package. Featurizers can be broadly categorized into composition-based, structure-based, and site-level descriptors.

Composition featurizers extract information from the elemental makeup of a material. For example, the *AtomicOrbitals* featurizer characterizes the orbital nature of the highest occupied molecular orbital (HOMO) and the lowest unoccupied molecular orbital (LUMO), whereas *AtomicPackingEfficiency* quantifies how efficiently atoms are packed in the structure. The *BandCenter* featurizer calculates the weighted average of the atomic orbital energy levels. *ElementFraction* and *ElementProperty* descriptors, the latter based on the Magpie dataset, represent elemental fractions and properties such as electronegativity, ionization energy, and atomic radius. *IonProperty* and *Miedema* featurizers predict ion formation tendencies and formation enthalpies, respectively. *Stoichiometry* encodes elemental ratios, whereas *TMetalFraction* computes the proportion of transition metals. *ValenceOrbital* describes the distribution of valence orbitals, and *YangSolidSolution* assesses the potential for solid-solution formation. Additionally, oxidation-aware descriptors include *ElectronegativityDiff*, which measures electronegativity differences, and *OxidationStates*, which encodes oxidation state information.

Structural featurizers focus on features derived from the spatial arrangement of atoms. *DensityFeatures* calculate the density and atomic density of the material, while *GlobalSymmetryFeatures* capture symmetry information, such as the crystal system and centrosymmetry. The *RadialDistributionFunction* describes atomic pair distributions at various distances, and both *CoulombMatrix* and *SineCoulombMatrix* encode electrostatic interactions in the structure. *EwaldEnergy* estimates the lattice energy using Ewald summation, and *BondFractions* quantify the proportion of specific bond types. *StructuralHeterogeneity* measures the variability in bond lengths and angles, while *MaximumPackingEfficiency* calculates the theoretical packing density. Other structural descriptors include *ChemicalOrdering*, which quantifies the degree of atomic ordering, and *XRDPowderPattern*, which simulates X-ray diffraction patterns.

Finally, site-level featurizers focus on local atomic environments. For instance, *AGNIFingerprints* generate atomic neighborhood fingerprints, whereas *AverageBondAngle* and *AverageBondLength* measure the mean bond angles and lengths based on Voronoi tesselation. *BondOrientationalParameter* captures the angular distribution of bonds, and *CoordinationNumber* quantifies the number of neighboring atoms. *CrystalNNFingerprint* encodes local atomic environments using crystal-graph techniques. Additional descriptors such as *GaussianSymmFunc* and *GeneralizedRadialDistributionFunction* represent atom-pair distributions and property-weighted distances, respectively. *LocalPropertyDifference* measures differences in properties between neighboring atoms, and fingerprints like *OPSiteFingerprint* and *VoronoiFingerprint* further characterize local atomic arrangements.

By leveraging this diverse set of features, the MatMiner featurizer enables a comprehensive representation of materials, facilitating accurate and interpretable machine learning predictions across various material properties. All featurizers are configured with default parameters tailored for broad applicability and optimized for MatMiner version 0.6.2.

# S2. Orbital field matrix featurizer

This study follows the original Orbital Field Matrix (OFM) implementation from Lam Pham et al. (2017), as also found in the MatMiner featurizer. The neutral valence shell electronic configurations of the elements can be represented as one-hot encoded vectors using an ordered dictionary, $D = \{s^1, s^2, p^1, p^2, ..., p^6, d^1, d^2, ..., d^{10}, f^1, f^2, ..., f^{14}\}$. For example, Na and Cl have the electronic configurations $[Ne]3s^1$ and $[Ne]3s^23p^5$, respectively. Sodium can then be represented by a one-hot encoded vector with position $s^1$ set to 1, while chlorine's vector has positions $s^2$ and $p^5$ set to 1 (the remaining entries are zeros). If we consider these elements within a crystal structure, as illustrated in Fig. S1, the OFM descriptor aims to capture the valence shell interactions at each site.

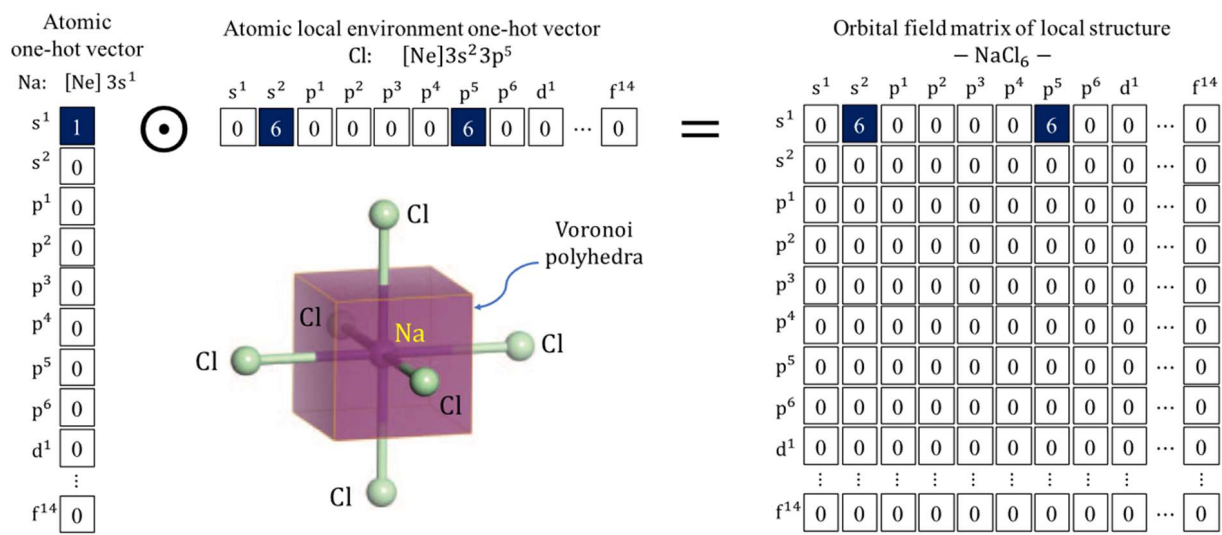

**Fig. S1** | OFM representation for a Na atom in a regular octahedral site surrounded by six Cl atoms. Sourced from Lam Pham et al.[1], reproduced with permission under the CC BY license.

It is important that the descriptor captures the site coordination and element distance from neighboring atoms. Therefore, the OFM for a central atom in a site ($X^p$) is defined as the weighted outer vector product of one-hot encoded atomic vectors, such as:

$$X_{ij}^p = \sum_{k=1}^{n_p} o_i^p o_j^k \frac{\theta_k^p}{\theta_{max}^p} \frac{1}{r_{pk}}. \qquad (1)$$

Here, $i, j \in D$, $k$ is the index of nearest-neighbor atoms, $n_p$ is the number of such atoms around site $p$, $\theta_k^p/\theta_{max}^p$ represents the weight of atom $k$ in the coordination of the central atom at site $p$, $\theta_k^p$ is the solid angle determined by the Voronoi polyhedron face separating $k$ and $p$, and $\theta_{max}^p$ is the maximum among $n_p$ them. $r_{pk}$ captures the distance separating atoms $p$ and $k$, also distinguishing elements with the same valence configuration. To construct the OFM for a crystal structure, local OFMs are summed, and the values are averaged by the number of sites:

$$F_{ij} = \frac{1}{N_p} \sum_p^{N_p} X_{ij}^p \qquad (2)$$

## S3. MEGNet framework and pretrained models

Fig. S2 illustrates the architecture of the MEGNet framework based on a graph convolutional network. As depicted in the figure, the final MLP of the model preceding the output contains two sequential dense layers of 32 and 16. These values can be tuned for hyperparameter optimization as elaborated in the next section, particularly the default architecture corresponds to $h_1 = 64$, $h_2 = 32$, and $h_3 = 16$. In which $h_1$ influences the MLPs inside the MEGNet blocks.

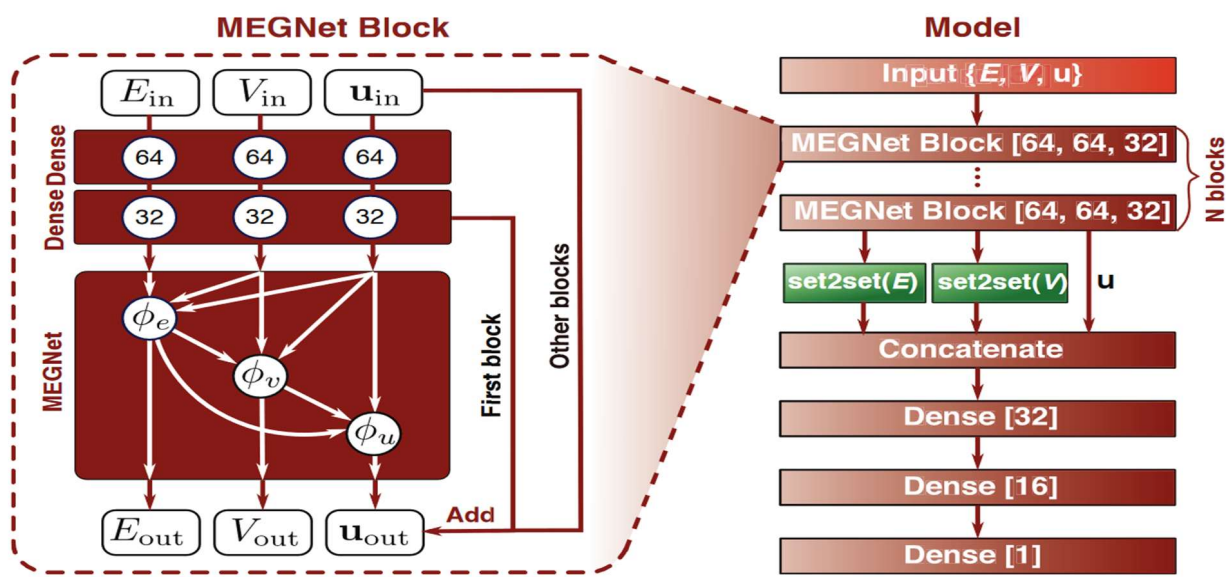

**Fig. S2** | Architecture of the MEGNet model. In the pretrained models used in this work, the same architecture was present with three MEGNet blocks. The numbers in brackets indicate the number of neurons for each layer. Reprinted (adapted) with permission from Chen et al[2]. Copyright 2024 American Chemical Society.

## S4. Latent-space optimization for OFM and MatMiner descriptors

An autoencoder is trained to compress the OFM features computed on the structures from Materials Project database, producing a latent-space representation that efficiently captures critical information from these structures. We consider a snapshot of the Materials Project database from 2019.04.02. This dataset excludes entries with a formation energy above 150 meV or those containing noble gases. Identical to the dataset used for band gap prediction task in MatBench. Different compression ratios (c.r.) are tested after hyperparameter tuning, the details of which are discussed in the following section. In Table S1, we compare two autoencoders with c.r. values of 20% and 10% by applying these compressed representations to replace the original OFM features for the new predictions. The 20% c.r. latent space notably improve the predictive accuracy over the original OFM features, likely due to transfer learning effects where chemical patterns from a broader dataset contribute to a more compact, chemically informative feature set. However, at 10% c.r., the compressed representation loses some chemical information, reducing its effectiveness compared to the original OFM features. Furthermore, as shown in Table S1, reducing the feature space to 20% c.r. using PCA is slightly less effective than using the autoencoder. Therefore, we retain the latent features from the OFM autoencoder, henceforth called ℓ-OFM for brevity.

**Table S1 | Mean absolute errors (MAE) for MODNet models on the matbench_perovskites task including pristine OFM features and different latent space reductions of OFM features in addition to the default MatMiner features. $n$ represents the number of features after removing constant features across the dataset. The shaded rows highlight the chosen latent-space representation using the autoencoder and the PCA-reduced representation with the same dimensions for comparison. In parentheses, the percentage MAE deviation is given with respect to the default MatMiner featurizer in MODNet.**

| Features | $n$ | MAE (eV) |
|---|---|---|
| Default MatMiner (MM) | 1020 | 0.0888 |
| MM + original OFM | 1020 + 943 | 0.0751 (−15.3%) |
| MM + latent OFM 20% c.r. (ℓ-OFM) | 1020 + 188 | 0.0743 (−16.2%) |
| MM + latent OFM 10% c.r. | 1020 + 94 | 0.0777 (−12.4%) |
| MM + PCA reduced OFM ($n = 188$) | 1020 + 188 | 0.0748 (−15.7%) |

Next, we benchmark latent-space representations of MatMiner features (ℓ-MM) against their original implementations (Table S2), evaluating their performance on the matbench_perovskites and matbench_mp_gap tasks. For the heat of formation predictions in perovskites, latent-space features consistently improve performance, which is attributed to transfer learning benefits from a larger dataset. For band gap predictions, a similar improvement is observed initially; however, compression beyond 60% c.r. leads to a decline in accuracy. This aligns with our hypothesis that the autoencoder aids transfer learning, which is comparable to the role of elemental embeddings in graph-based models[2].

Replacing the autoencoder with PCA resulted in a larger relative performance drop for the MatMiner features compared with OFM, likely due to the autoencoder's capacity to capture nonlinear patterns, which is essential for MatMiner's complex feature set. This result supports the choice of a 60% c.r. autoencoder for MatMiner features ($\ell$-MM), which offers an optimal balance of feature reduction with minimal accuracy loss, thereby favoring encoder-based over PCA-based latent-space representations.

**Table S2 | Evaluation of the effects of dimensionality reduction on default MatMiner features used on the MODNet model on the Matbench tasks matbench_perovskites and matbench_mp_gap. $n$ is the number of features (constant features across the dataset removed) for the respective model and N the number of samples comprised in the dataset. The percentage MAE deviation from the default MatMiner featurizer in MODNet is indicated between parentheses for each task.**

| Features used | Task | | | |
|---|---|---|---|---|
| | matbench perovskites (N=18,928) | | matbench mp_gap (N=106,113) | |
| | $n$ | MAE (eV) | $n$ | MAE (eV) |
| Default MatMiner | 1020 | 0.0888 ±0.0028 | 1264 | 0.2724 ±0.0052 |
| Latent MatMiner without compression (1:1 latent space) | 1264 | 0.0767 (−13.6%) | 1264 | 0.2542 (−6.7%) |
| Latent MatMiner 80% c.r. | 1011 | 0.0788 (−11.3%) | 1011 | 0.2809 (+3.1%) |
| Latent MatMiner 60% c.r. ($\ell$-MM) | 758 | 0.0793 (−10.7%) | 758 | 0.2911 (+6.8%) |
| Latent MatMiner 40% c.r. | 505 | 0.0844 (−4.9%) | 505 | 0.3280 (+20.4%) |
| PCA reduced MatMiner ($n = 758$) | 758 | 0.0816 (−8.1%) | 758 | 0.2968 (+8.9%) |

Finally, to streamline featurization, we implemented MEGNet GNN models as proxy featurizers for structure-based feature derivation. This approach allows these proxy GNNs, once trained, to be reused with new datasets, thereby significantly reducing the computational burden. The implementation details can be found in the section on hyperparameter tuning.

## S5. Descriptor-oriented GNNs for ℓ-OFM and ℓ-MM descriptors

In Table S3, the models including GNN-derived latent MatMiner features, ℓ-MM (via GNN), show an increase in MAE of 0.025 eV, most probably due to reconstruction errors. However, these models still outperform Automatminer and random forest benchmarks (Table 2) and allow faster featurization. For the OFM features, the GNN-derived latent representation performs nearly as well as the original, with only a 0.0051 eV decline. Combining latent features from both GNN models slightly reduces the MAE, highlighting the potential benefits of integrating multiple latent representations of chemical descriptors. We highlight that in the main paper ℓ-MM and ℓ-OFM are always obtained via GNN proxies, this distinction is only made in this section for evaluation.

These results emphasize the effectiveness of our proposed proxy GNN featurizers in capturing essential chemical information, even in the presence of reconstruction challenges, while also significantly reducing the computation time and making the feature-based models more efficient for large-scale applications. By further refining the models, such as training on larger, more carefully curated datasets, we can mitigate reconstruction errors and enhance the descriptor-oriented GNN featurizer in MatterVial to identify chemical patterns.

**Table S3 | Mean absolute errors (MAEs) for MODNet models on the matbench_perovskites task comparing the inclusion of latent features originally obtained from the autoencoder and through the GNN featurizers. The relative MAE deviation from the default MatMiner featurizer in MODNet is reported in parentheses.**

| Features | MAE (eV) |
|---|---|
| Default MatMiner (MM) | 0.0888 |
| MM + original OFM | 0.0751 (−15.3%) |
| $\ell$-MM | 0.0793 (−10.7%) |
| $\ell$-MM (via GNN) | 0.1052 (+18.5%) |
| MM + $\ell$-OFM | 0.0743 (−16.2%) |
| MM + $\ell$-OFM (via GNN) | 0.0794 (−10.6%) |
| $\ell$-MM (via GNN) + $\ell$-OFM (via GNN) | 0.0973 (+9.6%) |

## S6. Task-Oriented GNNs: investigating pretrained GNN feature integration via MVL featurizers

To incorporate pretrained GNN models from MVL as features, we extract the values from the last layers of the MLP regression head of the MEGNet model architecture. Table S4 presents a performance comparison for the matbench_perovskites task, incorporating the hidden layers with 32 neurons (referred to as MVL-32), the layers with 16 neurons (referred to as MVL-16), and both layers at once (MVL). Additionally, we conduct assessments on randomly selected subsets comprising 5000 samples and 1000 samples from the initial matbench_perovskites dataset to verify the consistency of our findings across smaller datasets and the effect of transfer learning.

Our analysis reveals a consistent enhancement in performance with the inclusion of the MVL-32 featurizer over the MVL-16 featurizer, irrespective of the dataset size. This improvement is attributed to a more general latent-space representation in the earlier layers of the model. When both layers are used concomitantly, the results are slightly better in general, which we attribute to MODNet capacity to wisely select the meaningful features. Notably, the percentage reduction in MAE compared with the exclusive use of MatMiner features increases as the dataset size decreases. This underscores the essence of transfer learning of this technique: transferring pre-acquired chemical knowledge from larger datasets to enhance performance on small datasets.

**Table S4 | Mean absolute errors (MAEs) for MODNet models on the matbench_perovskites task and subsets comparing the inclusion of features from pre-trained MEGNet models distributed by Materials Virtual Lab. *N* represents the size of the dataset used for the prediction. The relative MAE deviation from the default MatMiner featurizer in MODNet is reported in parentheses for each task.**

| Features | Task | | |
|---|---|---|---|
| | matbench perovskites (N=18,928) MAE (eV) | matbench perovskites (N=5,000) MAE (eV) | matbench perovskites (N=1,000) MAE (eV) |
| Default MatMiner (MM) | 0.0888 | 0.1667 | 0.2802 |
| MM + MVL-16 | 0.0752 (−15.3%) | 0.1202 (−27.9%) | 0.1862 (−33.5%) |
| MM + MVL-32 | 0.0726 (−18.2%) | 0.1167 (−30.0%) | 0.1749 (−37.6%) |
| MM + MVL | 0.0730 (−17.8%) | 0.1122 (−32.7%) | 0.1716 (−38.7%) |

# S7. Hyperparameter tuning for descriptor-oriented GNNs

## S7.1. Autoencoders' hyperparameters

The autoencoder architecture employed in this study consists of a feedforward neural network constructed with the Keras framework[4,] consisting of a single hidden layer for both the encoder and decoder. The number of neurons in the hidden layer is initialized at 2 times the number of features in the featurizer ($n$), whether OFM or general MatMiner features. Architectures with two hidden layers are excluded in the preliminary tests, as are hidden layers with a number of neurons smaller than $n$, which yielded poorer results. Hyperparameter tuning is conducted in two steps. Initially, the features' compression is fixed at 50% (approximately $n/2$ resulting features), and the optimal configuration is sought, considering the following possibilities, shown in Table S5. The Adam optimizer is utilized for weight optimization during backpropagation. For these combinations, the configurations with the smallest average reconstruction errors over three runs, employing a train-test split of 9:1, are presented in Table S6.

**Table S5 | Hyperparameters and corresponding values considered for the autoencoder optimization.**

| Hyperparameter | Possible Values |
| --- | --- |
| Batch Size | 16, 32, 64, 128 |
| Number of Epochs | 50, 100, 200, 300 |
| Learning Rate | 0.0005, 0.001, 0.002 |

**Table S6 | Best hyperparameters for autoencoders in this work, considering a 50% compression.**

| Encoded featurizer | Batch size | Number of epochs | Learning rate |
| --- | --- | --- | --- |
| OFM | 64 | 300 | 0.001 |
| MatMiner MODNet v.0.1.13 | 64 | 200 | 0.0005 |

Based on these parameters, we proceed with a similar approach to vary the number of neurons in the dense layer, ranging from $1.5n$ to $2.5n$ in increments of $0.1n$. This time, we test compressions of 20%, 50% and 80%. The combined loss for these compressions is assessed to identify the optimal architecture. As a result, the hidden layer sizes are determined to be $2.5n$ for the OFM featurizer and $2.2n$ for the MatMiner featurizer. The final architecture for each autoencoder is depicted in Fig. S3.

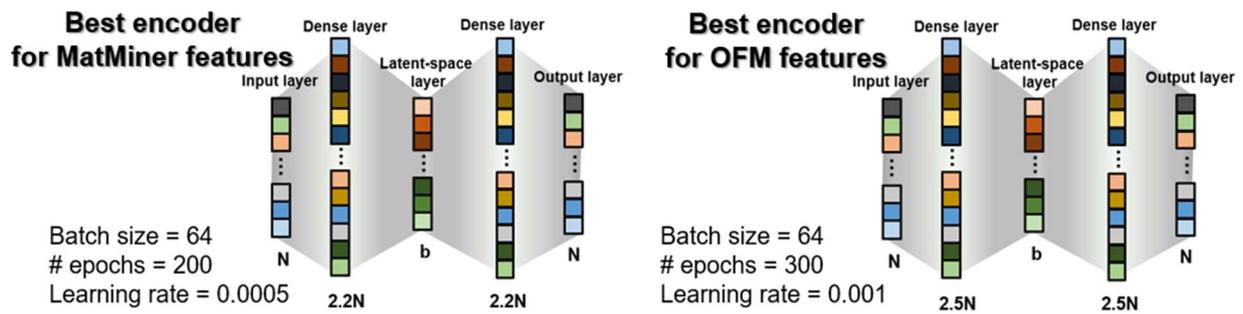

**Fig. S3** | Best autoencoder architectures found for MatMiner and OFM featurizers trained on matbench_v.0.1_mp_gap dataset.

Subsequently, the reconstruction loss is assessed for various levels of compression in each autoencoder, employing the same 9:1 train-test split. The results are outlined in Table S7 and Table S8. The encoder for MatMiner features consistently maintains the reconstruction error below 1%, even up to a compression to a latent-space size of 10% of the initial features. In the case of the OFM, the compression is highly efficient, remaining below 0.1% MAE for most of the tested latent-space sizes. Consequently, the reconstruction error is not anticipated to significantly impact the predictions. Nonetheless, the most suitable latent-space size must be determined by evaluating their performances in prediction tasks.

**Table S7 | Reconstruction errors with different compression ratios for the autoencoder for MODNet's v.0.1.13 MatMiner featurizer. Errors in data normalized to the interval 0 to 1, metric for losses is MSE.**

| Compression ratio | Latent $n$ | Train Loss | Validation Loss | Test MAE |
|---|---|---|---|---|
| *1.0** | 1264 | 7.91e-05 | 7.69e-05 | 0.004789 |
| *0.9* | 1137 | 8.66e-05 | 8.52e-05 | 0.005098 |
| *0.8* | 1011 | 8.59e-05 | 8.04e-05 | 0.005010 |
| *0.7* | 884 | 8.60e-05 | 9.20e-05 | 0.005309 |
| *0.6* | 758 | 9.27e-05 | 9.45e-05 | 0.005411 |
| *0.5* | 631 | 9.79e-05 | 1.06e-04 | 0.005733 |
| *0.45* | 568 | 1.02e-04 | 1.13e-04 | 0.005880 |
| *0.4* | 505 | 1.09e-04 | 1.14e-04 | 0.005929 |
| *0.35* | 442 | 1.14e-04 | 1.28e-04 | 0.006269 |
| *0.3* | 379 | 1.29e-04 | 1.44e-04 | 0.006624 |
| *0.25* | 316 | 1.53e-04 | 1.64e-04 | 0.006962 |
| *0.2* | 252 | 1.82e-04 | 1.85e-04 | 0.007387 |
| *0.15* | 189 | 2.38e-04 | 2.32e-04 | 0.008094 |
| *0.1* | 126 | 3.26e-04 | 3.24e-04 | 0.009452 |
| *0.05* | 63 | 5.92e-04 | 5.87e-04 | 0.012396 |

*\* A compression ratio of 1.0 indicates a remapping to a latent space with the same dimensions.*

*Note the number of dimensions may not precisely match the original featurizer's number of descriptors as some descriptors remain constant (0) throughout the dataset.*

**Table S8 | Reconstruction errors with different compression ratios for the autoencoder for OFM featurizer. Errors in data normalized to the interval from 0 to 1, the metric for losses is MSE.**

| Compression ratio | Latent $n$ | Train Loss | Validation Loss | Test MAE |
|---|---|---|---|---|
| *1.0** | 943 | 2.50e-05 | 3.26e-05 | 0.000898 |
| *0.9* | 848 | 1.45e-05 | 1.55e-05 | 0.000718 |
| *0.8* | 754 | 5.09e-06 | 6.69e-06 | 0.000534 |
| *0.7* | 660 | 3.80e-06 | 5.10e-06 | 0.000518 |
| *0.6* | 565 | 8.59e-06 | 1.04e-05 | 0.000915 |
| *0.5* | 471 | 3.51e-06 | 4.80e-06 | 0.000474 |
| *0.45* | 424 | 5.34e-06 | 6.53e-06 | 0.000507 |
| *0.4* | 377 | 7.25e-06 | 1.02e-05 | 0.000608 |
| *0.35* | 330 | 3.26e-06 | 5.01e-06 | 0.000442 |
| *0.3* | 282 | 4.82e-05 | 5.38e-05 | 0.001278 |
| *0.25* | 235 | 1.56e-05 | 1.61e-05 | 0.000750 |
| *0.2* | 188 | 4.70e-06 | 8.52e-06 | 0.000742 |
| *0.15* | 141 | 2.06e-05 | 2.66e-05 | 0.000790 |
| *0.1* | 94 | 1.45e-05 | 2.10e-05 | 0.000821 |
| *0.05* | 47 | 1.00e-05 | 1.13e-05 | 0.000837 |

*\* A compression ratio of 1.0 indicates a remapping to a latent space with the same dimensions. Note that the number of dimensions may not precisely match the original featurizer's number of descriptors as some descriptors remain constant (0) throughout the dataset.*

## S7.2. MEGNet models' hyperparameters

MEGNet models are trained to generate latent-space representations of encoded features (OFM and MatMiner features). In the case of the adjacent model, which produces general features based on the target property, both MEGNet and the coGN (Connectivity-Optimized Graph Network) model are utilized. No extensive hyperparameter tuning is performed for either graph neural network for adjacent models. The selected MEGNet and coGN parameters are detailed in Table S9.

**Table S9 | Hyperparameters applied to the adjacent MEGNet and coGN models. Parameters not referred to in the table follow the default values as of MEGNet's version 1.3.2 and coGN.**

| Hyperparameters MEGNet | Values | Hyperparameters coGN | Values |
|---|---|---|---|
| Number of blocks | 3 | Number of blocks | 4 |
| nfeat_bond | 100 | Embedding dimension | 64 |
| r_cutoff | 5 Å | r_cutoff | 5 Å |
| gaussian_width | 0.5 | Activation function | Swish |
| Number of epochs | 100 | Number of epochs | 800 |
| MLP architecture ($h_1 \times h_2 \times h_3$) | $64 \times 64 \times 128$ | MLP architecture ($h_1 \times h_2 \times h_3$) | $64 \times 64 \times 64$ |
| Batch size | 128 | Batch size | 64 |
| Learning Rate | 0.001 | Learning Rate | Polynomial decay from $5 \times 10^{-4}$ to $1 \times 10^{-5}$ |
| | | Number of neighbors (k) | 24 |

For the MEGNet models used to generate latent space features, hyperparameter tuning plays a crucial role. It is executed in three steps. Initially, the number of epochs varies across three different MLP architectures. Subsequently, the batch size (initially set at 32) and learning rate (default value of 0.001) are adjusted, with a new screening for the optimal number of epochs. Finally, a verification step is undertaken to assess whether increasing $h_1$ in the MLP architecture from 64 to 128 yields improvement. This process results in a total of 37 trained models, all evaluated on the same train-test split, with 20% of the dataset reserved for testing. All hyperparameter values considered for the respective optimization cases are presented in Table S10.

**Table S10 | Considered hyperparameter values for MEGNet models to generate encoded features for OFM and MatMiner featurizers.**

| Hyperparameter | | Possible Values |
|---|---|---|
| Number of epochs | | $10, 15, 20, 25, 30, 50, 70, 100$ |
| MLP architecture ($h_1 \times h_2 \times h_3$) | $h_1$ | $64, 128$ |
| | $h_2 \times h_3$ | $(16x32), (32x64), (64x128)$ |
| Batch size | | $16, 32, 64, 128$ |
| Learning Rate | | $0.0005, 0.001, 0.002$ |

A MEGNet model is trained to generate the latent OFM representation (20% compression), producing 188 features, and another MEGNet model to generate the latent representation of MatMiner features (60% compression), producing 758 features. A few selected results for both MEGNet models considered are shown in Table S11. We can observe the relevance of hyperparameter tuning on the final loss of these models. Despite the substantial number of features, the MEGNet framework is very successful in reproducing the latent space features directly from the structure. Even for the more heterogeneous and large set of MatMiner features, the error is about 0.03, which corresponds to 3% of the total variation within each normalized feature.

**Table S11 | MEGNet models' hyperparameters and reconstruction loss for generation of latent space features. Evaluation conducted on normalized features (range 0 to 1), highlighted in gray, was the best obtained model on the hyperparameter screening.**

| Encoded featurizer | Hyperparameters | | | | Reconstruction Loss (MAE) | |
|---|---|---|---|---|---|---|
| | Number of epochs | Batch size | Learning rate | MLP architecture ($h_1 x\ h_2 x\ h_3$) | Training | Test |
| Latent OFM, 20% compression (188 features) | 15 | 32 | 0.0005 | $64\ x\ 64\ x\ 32$ | 0.0180 | 0.0182 |
| | 25 | 64 | 0.001 | $64\ x\ 64\ x\ 32$ | 0.0164 | 0.0166 |
| | 15 | 128 | 0.001 | $64\ x\ 64\ x\ 32$ | 0.0137 | 0.0138 |
| | 25 | 32 | 0.0005 | $64\ x\ 128\ x\ 64$ | 0.0131 | 0.0132 |
| | 25 | 32 | 0.001 | $64\ x\ 128\ x\ 64$ | 0.0126 | 0.0127 |
| Latent MatMiner DeBreuck2020, 60% compression (758 features) | 50 | 16 | 0.001 | $64\ x\ 32\ x\ 16$ | 0.0671 | 0.0671 |
| | 20 | 64 | 0.0005 | $64\ x\ 128\ x\ 64$ | 0.0484 | 0.0486 |
| | 30 | 16 | 0.001 | $64\ x\ 32\ x\ 16$ | 0.0393 | 0.0393 |
| | 20 | 128 | 0.001 | $128\ x\ 128\ x\ 64$ | 0.0324 | 0.0326 |
| | 50 | 128 | 0.0005 | $64\ x\ 128\ x\ 64$ | 0.0306 | 0.0308 |

## S8. SISSO method in MatterVial

The Sure Independence Screening and Sparsifying Operator (SISSO) method[5] is an advanced symbolic regression technique designed to derive physically interpretable descriptors from an initially broad set of primary features. The workflow begins by generating a vast pool of candidate features through the recursive application of mathematical operators to fundamental descriptors extracted from materials data (e.g., via MatMiner). These operators include basic arithmetical functions (add, sub, mult, div), non-linear functions (sin, cos, exp, log), and specialized operations (e.g., abs_diff, square, cube, and root functions). These operations are arranged in a binary-expression-tree structure that respects physical constraints, such as unit consistency and valid operational domains (e.g., ensuring that arguments to logarithm functions remain positive).

However, the initial number of MatMiner features can be large, up to 1300 for structure-based tasks and 300 for composition-based, and the subsequent generation of candidate features becomes computationally prohibitive. To circumvent this computational challenge, we first reduce the pool of primary features to a more manageable set of 30. This is achieved using recursive feature elimination (RFE) guided by XGBoost models within a 5-fold cross-validation framework. In each iteration, features are marked for removal if they fall within the lowest 20th percentile of importance, and they are pruned from the set if at least three of the five models agree on their low rank. We term this integrated methodology xgb-rfe-SISSO, following the naming convention of similar hybrid approaches in the literature, like i-SISSO and rf-SISSO.

Once the candidate pool is established, the SISSO algorithm performs a sure-independence screening (SIS) step to rank features according to their individual correlations, typically quantified using Pearson coefficients, with the target material property. The top-ranked features are then refined through a $\ell_0$-norm based sparsification process, which constructs a minimal set of descriptors by selecting those symbols that not only possess high predictive power but also minimize redundancy. A distinctive advantage of the SISSO method is its ability to track multiple residuals from simpler, lower-dimensional models. This multi-residual approach enables the capture of independent and orthogonal contributions from candidate features, thereby enhancing both the robustness and interpretability of the final model.

In our implementation, the SISSO-derived symbolic expressions (as documented in the SIS_summary.txt file) include examples such as:

- ("TMetalFraction|transition metal fraction" + "ValenceOrbital|avg f valence electrons");
- ("ElectronegativityDiff|range EN difference" * "ElementFraction|O");
- (|"CrystalNNFingerprint|std_dev wt CN_2" - "ElementProperty|MagpieData minimum NValence"|);
- Etc.

These expressions are used to augment the original MatMiner feature set. Prior to the symbolic regression, the features are normalized using a robust scaler to facilitate the discovery of meaningful interactions. By merging these SISSO-based descriptors with the primary features, we obtain a rich and complementary feature space that synergistically enhances model performance. This approach, combining traditional descriptors with generated symbolic expressions, results in models that are not only more accurate but also offer valuable physical insights.

Below is an excerpt from our SISSO++ JSON configuration, which specifies all key hyperparameters and operational settings for the SISSO++ run used for most of the tasks:

```
{
  "data_file": "path_to_csv_with_MatMiner_features_and_target",
  "property_key": "target",
  "desc_dim": 2,
  "n_sis_select": 10,
  "max_rung": 1,
```

```
    "n_residual": 3,
    "calc_type": "regression",
    "min_abs_feat_val": 1e-05,
    "max_abs_feat_val": 100000000.0,
    "n_models_store": 1,
    "leave_out_frac": 0.05,
    "leave_out_inds": [],
    "opset": ["add", "sub", "abs_diff", "mult", "div", "inv", "abs", "exp", "log", "sin", "cos", "sq", "cb", "six_pow", "sqrt", "cbrt", "neg_exp"],
    "data_file_relative_to_json": true
}
```

This configuration not only sets the recursion and sparsification parameters but also carefully defines the operator set to balance the complexity and physical relevance of the generated descriptors. With this setup, a total of 20 formulas are generated during the SIS step in each task. These results form the basis for our enhanced descriptor space, which has been shown to improve material property prediction when integrated with our overall modeling framework.

Table S12 summarizes all the tasks (datasets) that were included to produce the SISSO_FORMULAS_v1 file:

**Table S12 | List of datasets and corresponding references used to generate rung 1 (pairs of features) SISSO formulas from MatMiner descriptors that are meaningful for materials predictions in diverse tasks**.

| Reference | Dataset Name |
|---|---|
| 6 | matbench_steels |
| 7 | matbench_jdft2d |
| 8 | matbench_phonons |
| 9 | matbench_expt_gap |
| 9 | matbench_expt_is_metal |
| 10,11 | matbench_glass |
| 12 | matbench_dielectric |
| 13 | matbench_perovskites |
| 14 | matbench_log_gvrh |

| Reference | Dataset Name |
|---|---|
| 14 | matbench_log_kvrh |
| 15 | matbench_mp_is_metal |
| 15 | matbench_mp_gap |
| 15 | matbench_mp_e_form |
| 16 | noemd_hse_pbe_diff |
| 16 | noemd_shg |

## S9. SHAP analysis definition and computation

In understanding complex machine learning models, SHAP (SHapley Additive exPlanations) emerges as a robust tool for revealing feature contributions[17]. SHAP values ($\phi$) provide a clear view of how each feature influences predictions, employing Shapley values from cooperative game theory obtained through the formula,

$$\phi_i(f) = \frac{1}{N} \sum_{S \subseteq N \setminus \{i\}} \frac{|S|!\,(|N| - |S| - 1)!}{|N|!} \left[f(S \cup \{i\}) - f(S)\right], \qquad (3)$$

ensures a fair distribution of contributions, capturing the unique impact of each feature on model predictions. In the equation, the factorial terms in the denominator are crucial for normalization. The factorial function, denoted by the exclamation mark, represents the product of all positive integers up to a given integer $n$. Specifically, the terms $|S|!\,(|N| - |S| - 1)!$ and $|N|!$ ensure that contributions from each feature are appropriately scaled relative to the size of subsets ($S$) and the total number of features ($N$). Normalization plays a pivotal role in ensuring a fair and unbiased distribution of feature contributions. By accounting for the varying sizes of feature subsets and the entire set of features, the formula effectively weighs each feature's contribution. This weighting ensures that the impact of individual features on model predictions is accurately reflected, without being overshadowed by the influence of larger feature sets.

All of our SHAP analysis calculations used the SHAP python library. For our MODNet models, using 300 samples and 500 perturbations on 24 CPU cores, gives converged feature importances and the analysis takes about 20 minutes to complete. In contrast, the XGBoost models used as surrogates for the latent GNN features are tree-based which allows for significantly faster SHAP calculations on the same configuration and hardware setup.

# S10. Interpretability of MODNet@MV(noORB) model for the matbench_perovskites task

Based on Figure 3 of the main paper, we give here a more descriptive analysis of the important features in the model separated by groups of features to highlight the synergy of these feature groups. Let's delve into the specific interpretations of the presented formulas. For instance, the MVL32_Eform_MP2019_#3 feature, derived from MVL's formation energy model, is associated with large electronegativity gaps (*ElectronegativityDiff|mean_EN_difference*), which promotes an ionic A/B-O/F bonding character. Its exponential term, $e^{-f_d}$ (where $f_d$ is *ValenceOrbital|frac_d_valence_electrons*), favors a low d-electron fraction, suggesting a preference for early transition metals. Similarly, MVL32_Eform_MP2019_#16 suggests that a negative band center (*BandCenter|band_center*), displayed by deep O-2p bands, for example, favors stability. Furthermore, alkaline-earth elements (with their higher number of s-valence electrons, from *ElementProperty|MagpieData_mean_NsValence*) coupled with a low $f_d$, are favored by increasing the feature value and, in turn, reducing the heat of formation.

SISSO rung 1 features such SISSO_matbench_glass_12 indicate a preference for structures with single early transition metals without lanthanides, based on the normalized $f_d$ and normal average deviation of f-valence electrons (*ElementProperty|MagpieData_avg_dev_NfValence*). This also correctly implies that lanthanides tend to prefer later transition metals in these structures. Beyond individual formulas, several SISSO features collectively highlight key factors. High melting point elements (*ElementProperty|MagpieData_mean_MeltingT*) combined with ordered CN=6 coordination (*CrystalNNFingerprint|std_dev_wt_CN_6*) lead to enhanced structural stability (seen in SISSO_matbench_log_kvrh_#12). An excess of d-electrons (*ValenceOrbital|avg_d_valence_electrons*) and heterogeneous unfilled d-states (*ElementProperty|MagpieData_avg_dev_NdUnfilled*) contribute to destabilizing electronic

effects (seen in SISSO_matbench_perovskites_13). Wide bandgap anions and compact early d-cations are optimal for achieving desirable ionic character in perovskites, this is suggested by the combined features *StructuralHeterogeneity|range_neighbor_distance_variation* and *DensityFeatures|density* in SISSO_matbench_mp_is_metal_#7. Lastly, SISSO_log_kvrh_#4, besides favoring dense packing also indicates that 3Å contacts are stabilizing factors for perovskites via the radial distribution function feature (*GeneralizedRDF|mean_Gaussian center=3.0 width=1.0*), deeming long bonds and poor packing as factors to increase the heat of formation, as expected.

The ROOST_Eform_PoolingLayer_#11 feature, from the ROOST model's pooling layer trained for the energy of formation, has a low $R^2$ of 0.20 however, we still see sensible behavior. It involves the minimum atomic number (*ElementProperty|MagpieData_minimum_Z*), stoichiometry, and the mean electronegativity (*ElementProperty|MagpieData_mean_Electronegativity*). Physically, a larger minimum atomic number often indicates heavier chemistries where the lightest atom in the lattice is relatively heavy, potentially leading to less strongly bound structures per atom. This feature thus encodes the idea that perovskites built from chemically diverse, heavier, and thermally mismatched elements tend to have higher heats of formation.

For brevity the features in the ℓ-OFM group, we obtained from the interpreter the most important OFM features by SHAP values, grouped them and indicated their proportionality to the feature value. These include specific orbital configurations and their proportionality to the feature value and, by consequence, to the heat of formation, they are as follows: $p^4$ - $p^3$(+), $p^3$ - $p^3$(+), $s^2$ - $s^1$(+), $p^5$ - $s^2$(+), $f^{14}$ - $s^2$(+), $d^5$ - $p^4$(-), and $s^2$ - $p^4$(-). As expected, the heat of formation of perovskites decreases when the interaction of orbitals $s^2$ and $p^4$ is present (OFM: $s^2$-$p^4$), characteristic of many oxide perovskites. Conversely, the presence of pnictogen elements (OFM: $p^3$-$p^3$, OFM: $p^4$-$p^3$) correlates with weaker chemical bonds, similar to halide perovskites

(OFM: p⁵-s²). Additionally, complete or almost complete d or f shells also correlate to a high heat of formation.

Finally, the physical interpretation of ℓ-MM_#730 feature for the heat of formation can be understood as a balance between destabilizing and stabilizing factors. Specifically, it suggests that a perovskite rich in wide-band-gap elements (*ElementProperty|MagpieData_mean_GSbandgap*) and containing structurally heterogeneous constituents (implied by the Gaussian symmetry function term) will tend to have a more positive heat of formation, thus reducing its stability. This aligns with observations that oxygen and fluorine, which have lower ground state band gaps among common anions (e.g., O (0 eV) < Br (1.457 eV) < F (1.97 eV) < Cl (2.493 eV) < N (6.437 eV) < I (6.456 eV)), are most frequent in perovskites with negative heats of formation. Conversely, the introduction of transition-metal species with many empty d states (*ElementProperty|MagpieData_maximum_NdUnfilled*) or elements with larger covalent radii (*ElementProperty|MagpieData_maximum_CovalentRadius*) offsets this penalty, decreasing $\Delta H_f$ and thereby favoring stability.

The varying $R^2$ values for the SISSO approximations also indicate how effectively complex GNN features can be represented by simpler, interpretable formulas. However, general MatMiner and OFM features still fall short in capturing chemically diverse local environments, highlighting the need for more meaningful, compact, and computationally inexpensive chemical descriptors. It's important to remember that this SHAP analysis is local; as MODNet leverages non-linear neural networks, features are not always utilized simultaneously. Interaction terms, for instance, can either amplify or diminish a feature's influence based on the range of another. Nevertheless, the model successfully incorporates established solid-state chemistry principles for perovskites, while also capturing subtle nuances that contribute to the enhanced accuracy achieved with deeper neural networks.

# S11. Interpretability of MODNet models compared

In this section, the SHAP value analysis plots are presented in Figs. S4–S6 for different MODNet models used to predict the heat of formation of perovskites (matbench_perovskites). These MODNet models incorporate increasingly complex and meaningful features, up to the model leveraging MatterVial + adjacent coGN features, which presents the lowest mean absolute error.

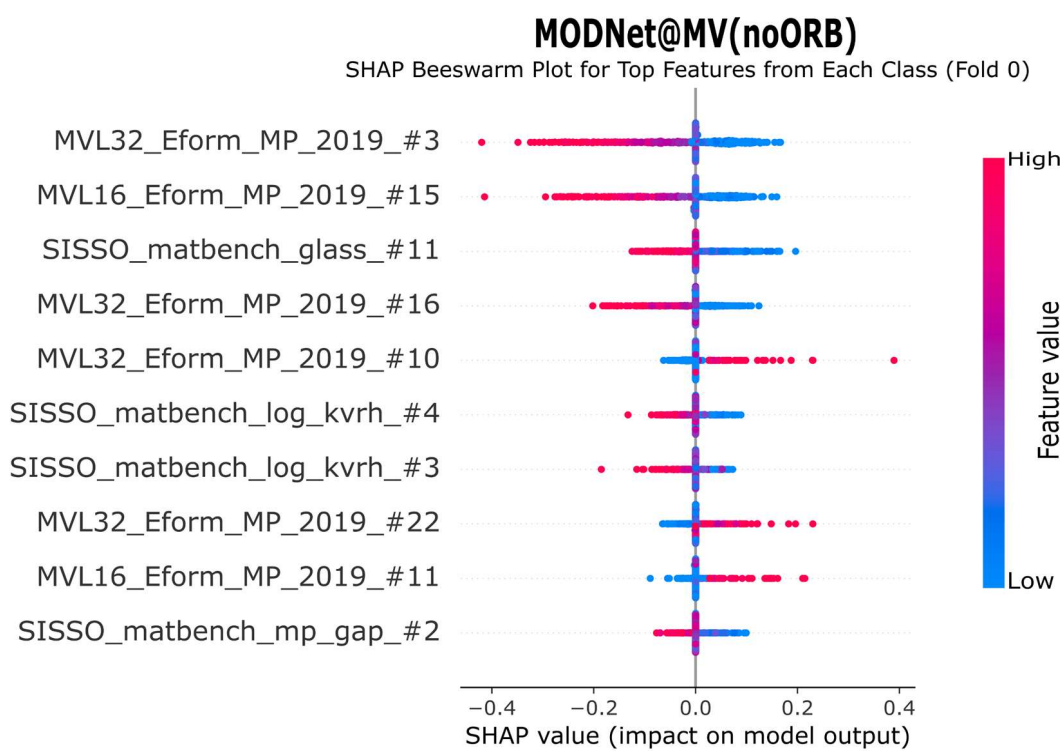

**Fig. S4** | SHAP analysis plot of the MODNet@MV(noORB) model, using MatterVial features excluding the ORB featurizer, on the matbench_perovskites task.

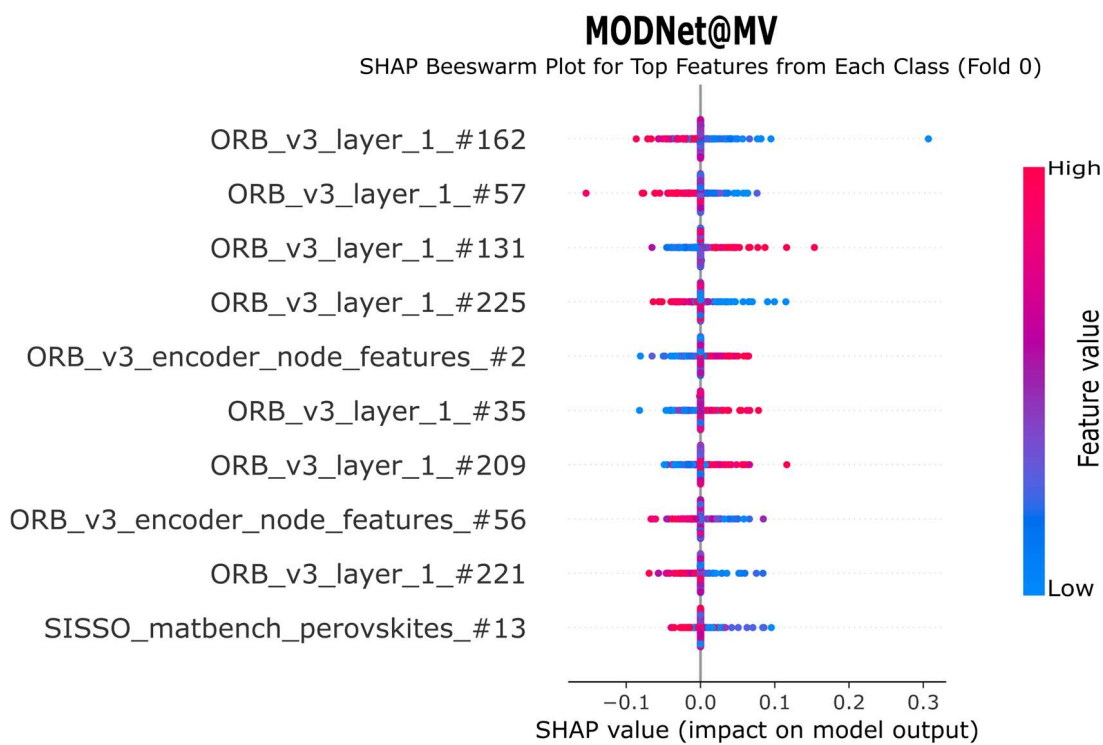

**Fig. S5** | SHAP analysis plot of the MODNet@MV model, using MatterVial features including the ORB featurizer, on the matbench_perovskites task.

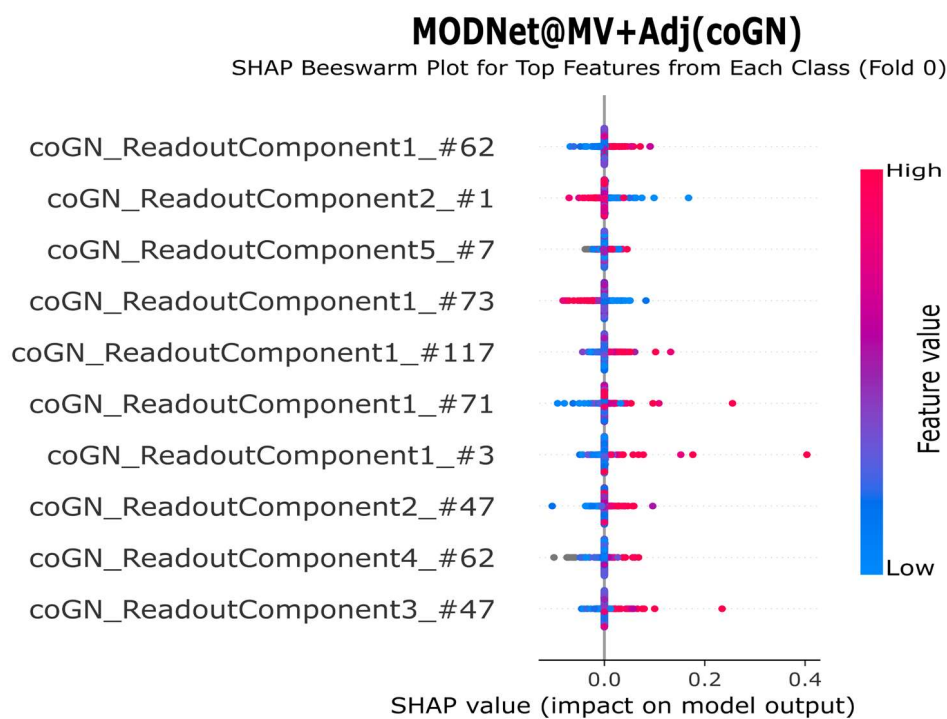

**Fig. S6** | SHAP analysis plot of the MODNet@MV+Adj(coGN) using MatterVial features and adjacent coGN model features, on the task of matbench_perovskites.

The coGN model, once trained for the matbench_perovskites task, is used to evaluate structures from the MP2018-stable dataset, which is fully featurized with interpretable

MatMiner descriptors and can be downloaded via MatterVial. Subsequently, we train surrogate XGBoost models that map the interpretable MatMiner and ℓ-OFM features to each adjacent-model feature. The top 30 features are passed to SISSO++ to predict an approximate formula. For the pretrained models included in MatterVial, the data are precomputed, and calling the interpreter module retrieves the approximate formulas.

In Fig. S7, we show the full SISSO formulas, with up to five terms, for the top features in each of the three analyzed models. We observe that the model with simpler features leverages primarily chemical information. When ORB features are introduced, the top feature leverages geometrical fingerprints and packing efficiency. The best-performing model seems to leverage the intertwined representations of chemical and geometrical features in the coGN features. Interestingly, the $R^2$ of the formula progressively decreases from the model without ORB and coGN to the model with the adjacent GNN, which can capture much more intricate relationships. This indicates that general chemical and geometrical descriptors become more limited as we move to models with deeper representations.

To further verify this observation, we analyzed the correlation of the top 20 features in each model with the interpretable MatMiner and OFM features instead of relying solely on the most important feature. These results are shown in Table S13, we can observe the same trend of more chemically oriented features in the model without ORB features, a bias towards using multiple geometrical features when the ORB features are introduced, and the resurge of more chemically oriented features in the model including full MatterVial featurizer with the addition of the adjacent coGN model features. Importantly, the correlation of the most important features decreases progressively, meaning that these interpretable descriptors become more and more limited to represent the nuances that the GNN features learn.

**MODNet@MV(noORB):**   **MVL32_Eform_MP_2019_#3**, $R_{form.}^2 = 0.91$

$$-5.4\text{e-}02 - 4.69 \times ((ElementProperty|MagpieData\_mode\_CovalentRadius \times ValenceOrbital|frac\_p\_valence\_electrons)$$
$$\times |ElementProperty|MagpieData\_mean\_GSbandgap - ValenceOrbital|frac\_d\_valence\_electrons|)$$
$$- 0.94 \times \left(\frac{|ElementProperty|MagpieData\_mean\_NpValence - ElementProperty|MagpieData\_minimum\_Column|}{ElementProperty|MagpieData\_minimum\_Column + IonProperty|max\_ionic\_char}\right)$$
$$+ 0.65 \times \left((ElementProperty|MagpieData\_avg\_dev\_NdValence \times Miedema|Miedema\_deltaH\_inter) - \sqrt[3]{\ell\text{-}OFM\_\#99}\right)$$
$$- 8.10 \times ((ElementProperty|MagpieData\_mean\_NpValence \times ElementProperty|MagpieData\_avg\_dev\_NUnfilled)$$
$$-(\ell\text{-}OFM\_\#79 \times ElectronegativityDiff|minimum\_EN\_difference))$$
$$+ 10.44 \times (|ElementProperty|MagpieData\_mean\_NpValence - ValenceOrbital|frac\_d\_valence\_electrons|$$
$$\times (ValenceOrbital|frac\_d\_valence\_electrons - ElementProperty|MagpieData\_avg\_dev\_Electronegativity))$$

**MODNet@MV(ORB):**   **ORB_v3_layer_1_#162**, $R_{form.}^2 = 0.84$

$$-1.32 + 8364.36 \times |(GeneralizedRDF|mean\_Gaussian\_center = 2.0\_width = 1.0 \times ElementProperty|MagpieData\_mean\_GSvolume\_pa)$$
$$- |DensityFeatures|packing\_fraction - MaximumPackingEfficiency|max\_packing\_efficiency||$$
$$- 9.9e-3 \times \left(\frac{TMetalFraction|transition\_metal\_fraction/ElementProperty|MagpieData\_maximum\_MeltingT}{ElementProperty|MagpieData\_avg\_dev\_NValence + DensityFeatures|packing\_fraction}\right)$$
$$- 1.34 \times ((GeneralizedRDF|mean\_Gaussian\_center = 2.0\_width = 1.0 \times ElementProperty|MagpieData\_mean\_MeltingT)$$
$$-(ElementProperty|MagpieData\_mean\_CovalentRadius \times BandCenter|band\_center))$$
$$+ 2.02 \times \left|(ElementProperty|MagpieData\_mean\_NdUnfilled \times ElementFraction|O) - \sqrt[3]{ElementProperty|MagpieData\_mean\_NpUnfilled}\right|$$
$$+ 2.35 \times |(GeneralizedRDF|mean\_Gaussian\_center = 2.0\_width = 1.0 \times ElementProperty|MagpieData\_mean\_GSvolume\_pa)$$
$$- (ElementFraction|O + ElementProperty|MagpieData\_mean\_Electronegativity)|$$

**MODNet@MV+adj(coGN):**   **coGN_ReadoutComponent1_#62**, $R_{form.}^2 = 0.70$

$$0.72 + 0.47 \times |(ElementProperty|MagpieData\_minimum\_GSvolume\_pa \times CrystalNNFingerprint|std\_dev\_linear\_CN\_2)$$
$$-(ElementProperty|MagpieData\_avg\_dev\_NUnfilled \times ElementProperty|MagpieData\_mean\_NdValence)|$$
$$+ 2.1e-5 \times \left(\frac{ElementProperty|MagpieData\_mean\_NdValence/DensityFeatures|packing\_fraction}{|MaximumPackingEfficiency|max\_packing\_efficiency - ElementProperty|MagpieData\_mean\_NdValence|}\right)$$
$$+ 6.5e-2 \times \left(\frac{CoulombMatrix|coulomb\_matrix\_eig\_1 \times ElementProperty|MagpieData\_avg\_dev\_NUnfilled}{ElementProperty|MagpieData\_minimum\_Electronegativity - ElementProperty|MagpieData\_mean\_NdUnfilled}\right)$$
$$- 4.6e-3 \times \left(\frac{|ElementProperty|MagpieData\_mean\_NdUnfilled - CrystalNNFingerprint|mean\_linear\_CN\_2|}{|CrystalNNFingerprint|mean\_octahedral\_CN\_6 - DensityFeatures|packing\_fraction|}\right)$$
$$- 1.67 \times \left((ValenceOrbital|frac\_d\_valence\_electrons - CrystalNNFingerprint|mean\_linear\_CN\_2) \times \left(\frac{ElementProperty|MagpieData\_maximum\_MeltingT}{ElementProperty|MagpieData\_minimum\_Electronegativity}\right)\right)$$

**Fig. S7** | SISSO formulas retrieved with the MatterVial interpreter module for the top feature in each of the analyzed MODNet models for the matbench_perovskites task. Deep GNN features from MVL, ORB, or coGN, were approximated with MatMiner features from the *DeBreuck2020Featurizer* and ℓ-OFM features, the coefficient of determination of the formulas against the real features is shown.

**Table S13 | Correlation between MatMiner and OFM of the top 20 features in the three analyzed MODNet models for matbench_perovskites.**

| MODNet@MV(noORB) | Corr. | MODNet@MV | Corr. | MODNet@MV+adj(coGN) | Corr. |
|---|---|---|---|---|---|
| ElementFraction\|N | 0.3482 | ElementProperty\|MagpieData_mean_GSvolume_pa | 0.2419 | ElementFraction\|N | 0.1967 |
| ElementProperty\|MagpieData_mean_SpaceGroupNumber | 0.3392 | ElementProperty\|MagpieData_maximum_GSvolume_pa | 0.2408 | ElementProperty\|MagpieData_mean_SpaceGroupNumber | 0.1966 |
| LocalPropertyDifference\|mean_local_diff_in_Electronegativity | 0.3389 | ElementProperty\|MagpieData_avg_dev_GSvolume_pa | 0.2395 | ElementProperty\|MagpieData_mean_GSbandgap | 0.1936 |
| ElementProperty\|MagpieData_mean_GSbandgap | 0.3328 | BandCenter\|band_center | 0.2386 | OFM:s²_-_p³ | 0.1906 |
| IonProperty\|avg_ionic_char | 0.3328 | ElementProperty\|MagpieData_range_GSvolume_pa | 0.2385 | OFM:p³_-_s² | 0.19 |
| ElementProperty\|MagpieData_range_Electronegativity | 0.3327 | ElementProperty\|MagpieData_minimum_Electronegativity | 0.2331 | ElementProperty\|MagpieData_range_MeltingT | 0.1895 |
| IonProperty\|max_ionic_char | 0.3278 | ElementProperty\|MagpieData_maximum_CovalentRadius | 0.2319 | ElementProperty\|MagpieData_maximum_MeltingT | 0.1895 |
| ElementProperty\|MagpieData_avg_dev_Electronegativity | 0.325 | ElementProperty\|MagpieData_mean_Electronegativity | 0.2295 | ElementProperty\|MagpieData_avg_dev_MeltingT | 0.1883 |
| ElementProperty\|MagpieData_minimum_MeltingT | 0.3237 | ElementProperty\|MagpieData_minimum_MendeleevNumber | 0.2275 | ElementProperty\|MagpieData_avg_dev_SpaceGroupNumber | 0.1851 |
| ElementProperty\|MagpieData_mode_GSbandgap | 0.3178 | ElementProperty\|MagpieData_mean_CovalentRadius | 0.2271 | ElementProperty\|MagpieData_minimum_MeltingT | 0.1817 |
| ElementProperty\|MagpieData_maximum_Column | 0.3173 | ElementProperty\|MagpieData_range_CovalentRadius | 0.226 | ElementProperty\|MagpieData_mean_MeltingT | 0.1817 |
| ElementProperty\|MagpieData_maximum_NpValence | 0.3173 | ElementProperty\|MagpieData_avg_dev_MendeleevNumber | 0.2246 | ElementProperty\|MagpieData_avg_dev_GSbandgap | 0.179 |
| ElementProperty\|MagpieData_mode_SpaceGroupNumber | 0.3165 | ElementProperty\|MagpieData_mean_MendeleevNumber | 0.2222 | YangSolidSolution\|Yang_omega | 0.1783 |
| ElementProperty\|MagpieData_avg_dev_SpaceGroupNumber | 0.3136 | ElementProperty\|MagpieData_range_MendeleevNumber | 0.2214 | OFM:p³_-_p³ | 0.1767 |
| ElementProperty\|MagpieData_maximum_Electronegativity | 0.3103 | IonProperty\|avg_ionic_char | 0.2172 | ElementProperty\|MagpieData_mode_GSbandgap | 0.1767 |
| ElementProperty\|MagpieData_maximum_MendeleevNumber | 0.3078 | ElementProperty\|MagpieData_minimum_NValence | 0.2154 | ElementProperty\|MagpieData_maximum_Column | 0.1765 |
| ElementProperty\|MagpieData_minimum_CovalentRadius | 0.2992 | AGNIFingerPrint\|std_dev_AGNI_eta=2_89e+00 | 0.2151 | ElementProperty\|MagpieData_maximum_NpValence | 0.1765 |
| ElementProperty\|MagpieData_minimum_SpaceGroupNumber | 0.298 | ElementProperty\|MagpieData_avg_dev_CovalentRadius | 0.2143 | ElementFraction\|O | 0.1763 |
| OFM:p³_-_s² | 0.2964 | GaussianSymmFunc\|std_dev_G2_4.0 | 0.2143 | ElementProperty\|MagpieData_mode_SpaceGroupNumber | 0.176 |
| OFM:s²_-_p³ | 0.2958 | ElementProperty\|MagpieData_avg_dev_Electronegativity | 0.214 | ElementProperty\|MagpieData_avg_dev_NpValence | 0.1758 |
| ElementProperty\|MagpieData_avg_dev_GSbandgap | 0.2934 | VoronoiFingerprint\|mean_Voro_dist_minimum | 0.2115 | ElementProperty\|MagpieData_maximum_Electronegativity | 0.1757 |
| SineCoulombMatrix\|sine_coulomb_matrix_eig_3 | 0.2915 | LocalPropertyDifference\|mean_local_diff_in_Electronegativity | 0.2113 | OFM:f⁴_-_d¹⁰ | 0.1753 |
| CoulombMatrix\|coulomb_matrix_eig_3 | 0.2889 | AverageBondLength\|mean_Average_bond_length | 0.21 | ElementProperty\|MagpieData_mean_Electronegativity | 0.1743 |
| ElementProperty\|MagpieData_mean_NpValence | 0.2866 | ElementProperty\|MagpieData_range_Electronegativity | 0.2096 | IonProperty\|avg_ionic_char | 0.1729 |
| ValenceOrbital\|avg_p_valence_electrons | 0.2866 | IonProperty\|max_ionic_char | 0.2078 | ElementProperty\|MagpieData_mean_NpValence | 0.1716 |
| ElementProperty\|MagpieData_range_SpaceGroupNumber | 0.2849 | VoronoiFingerprint\|std_dev_Voro_vol_sum | 0.2074 | ValenceOrbital\|avg_p_valence_electrons | 0.1716 |
| ElementProperty\|MagpieData_mode_GSvolume_pa | 0.2807 | DensityFeatures\|packing_fraction | 0.2042 | ElementProperty\|MagpieData_maximum_MendeleevNumber | 0.1706 |
| Miedema\|Miedema_deltaH_amor | 0.2801 | DensityFeatures\|density | 0.2037 | VoronoiFingerprint\|mean_Voro_dist_maximum | 0.1705 |
| OFM:f⁴_-_d¹⁰ | 0.276 | LocalPropertyDifference\|std_dev_diff_in_Electronegativity | 0.2018 | ElementProperty\|MagpieData_avg_dev_Electronegativity | 0.1703 |
| ElementProperty\|MagpieData_range_NpValence | 0.2756 | AGNIFingerPrint\|std_dev_AGNI_eta=4_43e+00 | 0.1996 | ElementProperty\|MagpieData_minimum_CovalentRadius | 0.1697 |
| OFM:p³_-_p³ | 0.2705 | VoronoiFingerprint\|mean_Voro_vol_sum | 0.1976 | ElementProperty\|MagpieData_range_Electronegativity | 0.1697 |
| ElementFraction\|O | 0.2695 | DensityFeatures\|vpa | 0.1972 | LocalPropertyDifference\|mean_local_difference_in_Electronegativity | 0.1695 |
| Miedema\|Miedema_deltaH_inter | 0.2663 | ElementProperty\|MagpieData_mean_MeltingT | 0.1971 | IonProperty\|max_ionic_char | 0.1686 |
| SineCoulombMatrix\|sine_coulomb_matrix_eig_4 | 0.2627 | ElementProperty\|MagpieData_avg_dev_MeltingT | 0.1963 | ElementProperty\|MagpieData_minimum_SpaceGroupNumber | 0.1668 |
| ElementProperty\|MagpieData_range_MendeleevNumber | 0.2601 | ElementProperty\|MagpieData_mean_Column | 0.1948 | ElementProperty\|MagpieData_range_SpaceGroupNumber | 0.1638 |
| AtomicPackingEfficiency\|dist_from_3_clusters__APE__<_0.010 | 0.2575 | AverageBondLength\|std_dev_Average_bond_length | 0.194 | BandCenter\|band_center | 0.1629 |
| CoulombMatrix\|coulomb_matrix_eig_4 | 0.2571 | MaximumPackingEfficiency\|max_packing_efficiency | 0.1927 | CrystalNNFingerprint\|mean_linear_CN_2 | 0.1626 |
| AtomicPackingEfficiency\|dist_from_1_clusters__APE__<_0.010 | 0.2568 | ElementProperty\|MagpieData_mean_NdValence | 0.1926 | VoronoiFingerprint\|mean_Voro_dist_minimum | 0.1626 |
| ElementProperty\|MagpieData_minimum_Electronegativity | 0.2567 | ValenceOrbital\|avg_d_valence_electrons | 0.1926 | OFM:p³_-_d¹⁰ | 0.1623 |
| ElementProperty\|MagpieData_avg_dev_MendeleevNumber | 0.2562 | GeneralizedRDF\|std_dev_Gaussian_center=2.0_width=1.0 | 0.1921 | OFM:d¹⁰_-_p³ | 0.1623 |
| OFM:p⁴_-_s² | 0.2561 | VoronoiFingerprint\|mean_Voro_area_sum | 0.1912 | VoronoiFingerprint\|mean_Voro_dist_mean | 0.1622 |
| ElementFraction\|F | 0.256 | ValenceOrbital\|frac_s_valence_electrons | 0.1908 | ElementProperty\|MagpieData_range_NpValence | 0.1618 |
| OFM:s²_-_p⁴ | 0.2547 | ElementProperty\|MagpieData_avg_dev_Column | 0.1897 | OFM:f¹⁰_-_d³ | 0.1599 |
| ElementProperty\|MagpieData_minimum_GSvolume_pa | 0.2545 | GeneralizedRDF\|std_dev_Gaussian_center=4.0_width=1.0 | 0.1893 | OFM:d¹⁰_-_f⁴ | 0.1594 |
| AtomicPackingEfficiency\|dist_from_5_clusters__APE____0_010 | 0.2536 | CrystalNNFingerprint\|mean_linear_CN_2 | 0.1886 | CoulombMatrix\|coulomb_matrix_eig_3 | 0.1594 |
| ValenceOrbital\|frac_p_valence_electrons | 0.2498 | StructuralHeterogeneity\|mean_neighbor_distance_variation | 0.1881 | GlobalSymmetryFeatures\|crystal_system_int | 0.159 |
| LocalPropertyDifference\|std_dev_diff_in_Electronegativity | 0.248 | ElementProperty\|MagpieData_avg_dev_NdValence | 0.1879 | OFM:d³_-_f¹⁰ | 0.1584 |
| ElementProperty\|MagpieData_range_CovalentRadius | 0.2477 | ElementProperty\|MagpieData_minimum_Column | 0.1878 | ElementProperty\|MagpieData_mode_GSvolume_pa | 0.1581 |
| ElementProperty\|MagpieData_mode_NUnfilled | 0.2455 | YangSolidSolution\|Yang_delta | 0.1873 | AverageBondLength\|mean_Average_bond_length | 0.1578 |
| OFM:d⁸_-_f⁷ | 0.2453 | VoronoiFingerprint\|mean_Voro_vol_mean | 0.1873 | CrystalNNFingerprint\|mean_wt_CN_2 | 0.1574 |

# Supplementary Information References